\documentclass[onecolumn,journal]{IEEEtran}

\makeatletter\def\input@path{{current}}\makeatother

\usepackage{amsmath,amsfonts,amsthm}
\usepackage{algorithmic}
\usepackage{graphicx}
\usepackage{textcomp}
\usepackage{color}
\usepackage{xcolor}
\usepackage[utf8]{inputenc}
\usepackage[T1]{fontenc}

\usepackage{xfrac}

\usepackage[nolist]{acronym}

\usepackage{array}
\newcommand{\PreserveBackslash}[1]{\let\temp=\\#1\let\\=\temp}
\newcolumntype{C}[1]{>{\PreserveBackslash\centering}p{#1}}
\newcolumntype{R}[1]{>{\PreserveBackslash\raggedleft}p{#1}}
\newcolumntype{L}[1]{>{\PreserveBackslash\raggedright}p{#1}}

\usepackage{subcaption}

\def\BibTeX{{\rm B\kern-.05em{\sc i\kern-.025em b}\kern-.08em
    T\kern-.1667em\lower.7ex\hbox{E}\kern-.125emX}}

\newcommand{\blue}[1]{#1} 

\newcommand{\blueeffect}[1]{#1} 

\newcommand{\stkout}[1]{\ifmmode\text{\sout{\ensuremath{#1}}}\else\sout{#1}\fi}

\newcommand\reallyremove[1]{{}}
\definecolor{impactfulgreen}{rgb}{0.0 0.5 0.00}
\definecolor{backgroundgreen}{rgb}{0.0 0.7 0.0}
\definecolor{olivegreen}{rgb}{0.0 0.8 0.2}
\definecolor{darkblue}{rgb}{0 0 0.7}

\newcommand{\expminus}{e^{-\int_0^t \Qware(p) \, dp }}
\newcommand{\expplus}{e^{\int_0^\tau \Qware(p) \, dp }}
\newcommand{\bintegral}{\Fnominal \int_0^t \expplus \bonware(\tau) \, d \tau}  

\newcommand{\functionality}{{{F}}}

\newcommand\bonware{\mathcal{B}}
\newcommand\malware{\mathcal{M}}
\newcommand\Qware{\mathcal{Q}}
\newcommand\ma{\mathcal{A}} 

\newcommand\Fnominal{\blue{F_\text{N}}{}}

\DeclareMathOperator{\erf}{erf}

\newcommand\impact{\blueeffect{effectiveness}} 
\newcommand\impactfulness{\blueeffect{effectiveness}}
\newcommand\impacts{\blueeffect{effectivenesses}{}}

\IEEEoverridecommandlockouts
\begin{document}

\graphicspath{{./figures/}}

\newcommand\Description[1]{}

\title{Quantitative Measurement of Cyber Resilience: Modeling and Experimentation\thanks{\thankyou}}

\newcommand\arlsuper{$^1$}
\newcommand\ucisuper{$^4$}
\newcommand\psusuper{$^2$}
\newcommand\icfsuper{$^3$}

\newcommand\listauthors{Michael J. Weisman\arlsuper, Alexander
Kott\arlsuper, Jason
E. Ellis\arlsuper, Brian J. Murphy\psusuper, \\Travis
W. Parker\icfsuper, Sidney Smith\arlsuper, Joachim Vandekerckhove\ucisuper}

\newcommand\arl{\noindent \arlsuper \textit{DEVCOM Army Research
  Laboratory} \\}

\newcommand\uci{\ucisuper \textit{University of California, Irvine} \\}

\newcommand\icf{\icfsuper \textit{ICF International} \\}

\newcommand\psu{\psusuper \textit{Pennsylvania State University} \\}

\newcommand\thankyou{\hspace{-4mm} \ousd \coopagree \locations}

\newcommand\ousd{This work was partially funded by Cyber Technologies,
    Deputy CTO for Critical Technologies/Applied Technology, Office of
    the Under Secretary of Defense Research and Engineering.\\ \\}

\newcommand\coopagree{\ucisuper Research was sponsored by the Army Research
Laboratory and was accomplished under Cooperative Agreement Number
W911NF-21-2-0284. JV was additionally supported by NSF \#1850849 and \#2051186.\\

\hspace{-4.5mm} The views and conclusions contained in this
document are those of the authors and should not be interpreted as
representing the official policies, either expressed or implied, of
the Army Research Laboratory or the U.S. Government. The
U.S. Government is authorized to reproduce and distribute reprints for
Government purposes notwithstanding any copyright notation herein.\\ \\}

\newcommand\locations{\arl \psu \icf \uci }

\newcommand\nobody{\IEEEauthorblockN{\phantom{nobody}\\\phantom{nowhere}\\}}

\author{\listauthors}

\maketitle
\thispagestyle{plain}
\textbf{Distribution Statement A:} Approved for public release; distribution is
unlimited.
\\

\begin{abstract}
    \label{abstract}
\textbf{Abstract} Cyber resilience is the ability of a system to resist and recover from a cyber attack, thereby    restoring the system’s functionality. Effective design and development of a cyber resilient system requires experimental methods and tools for quantitative measuring of cyber resilience. This paper describes an experimental method and test bed for obtaining resilience-relevant data as a system (in our case -- a truck) traverses its route, in repeatable, systematic experiments. We model a truck equipped with an autonomous cyber-defense system and which also includes inherent physical resilience features. When attacked by malware, this ensemble of cyber-physical features (i.e., ``bonware'') strives to resist and recover from the performance degradation caused by the malware’s attack. We propose parsimonious mathematical models to aid in quantifying systems' resilience to cyber attacks. Using the models, we identify quantitative characteristics obtainable from experimental data, and show that these characteristics can serve as useful quantitative measures of cyber resilience.
\end{abstract}

\section{Introduction}
Resilience continues to gain attention as a key property of cyber and cyber-physical systems, for the purposes of cyber defense. Although definitions vary, it is generally agreed that cyber resilience refers to the ability of a system to resist and recover from a cyber compromise that degrades the performance of the system \cite{kott2022, weisman23, kott2019cyber}. One way to conceptualize resilience is as the ability of a system to absorb stress elastically and return to the original functionality once the stress is removed or nullified \cite{smith23towards}. Resilience should not be conflated with risk or security \cite{linkov2018risk}.

To make the discussion more concrete, consider the example of a truck which attempts to complete its goal of delivering heavy cargo. The cyber adversary's malware successfully gains access to the Controller Area Network (CAN bus) of the truck \cite{bozdal2018august}. Then, the malware executes cyber attacks by sending a combination of messages intended to degrade the truck's performance and diminish its ability to complete its goal. We assume that the malware is at least partly successful, and the truck indeed begins to experience a degradation of its goal-relevant performance.

At this point, we expect the truck's resilience-relevant elements to resist the degradation and then to recover its performance to a satisfactory level, within an acceptably short time period. These ``resilience-relevant elements'' might be of several kinds. First, because the truck is a cyber-physical system, certain physical characteristics of the truck's mechanisms will provide a degree of resilience.  For example, the cooling system of an internal combustion engine will exhibit a significant resistance to overheating even if the malware succeeds in misrepresenting the temperature sensors data, especially with a passive cooling system and with mechanical thermostats that do not depend on electronic manipulation of sensor data \cite{Pang2004}.
Second, appropriate defensive software residing on the truck continually monitors and analyzes the information passing through the CAN bus \cite{kott2018}. When the situation appears suspicious, it may take actions such as blocking or correcting potentially malicious messages. Third, it is possible that a remote monitoring center, staffed with experienced human cyber defenders, will detect a cyber compromise and will provide corrective actions remotely \cite{kott2021cyber}.

For the purposes of this paper, we assume that the remote monitoring and resilience via external intervention is impossible \cite{kott2020doers}. This may be the case if the truck cannot use radio communications due to environmental constraints (e.g., operating in a remote mountainous area), or if the malware spoofs or blocks communication channels of the truck. Therefore, in this paper we assume that resilience is provided by the first two classes of resilience-relevant elements. Here, by analogy with malware, we call these ``bonware'' -- a combination of physical and cyber features of the truck that serve to resist and recover from a cyber compromise.

A key challenge in the field of cyber resilience is quantifying or measuring resilience. Indeed, no engineering discipline achieved significant maturity without being able to measure the properties of phenomena relevant to the discipline \cite{kott2021cyber}. Developers of systems like a truck must be able to quantify the resilience of the truck under development in order to know whether the features they introduce in the truck improve its cyber resilience, or make it worse. Similarly, buyers of the truck need to know how to specify quantitatively the resilience of the truck, and how to test resilience quantitatively in order to determine whether the product meets their specifications. 

In this paper, we report results of a project called \textit{Quantitative Measurement of Cyber Resilience} (QMoCR) in which our research team seeks to identify quantitative characteristics of systems' responses to cyber compromises that can be derived from repeatable, systematic experiments. Briefly, we have constructed a test-bed in which a surrogate truck is subjected to controlled cyber attacks produced by malware. The truck is equipped with an autonomous cyber-defense system \cite{kott2018,kott2020doers} and also has some inherent physical resilience features. This ensemble of cyber-physical features (i.e., bonware) strives to resist and recover from the performance degradation caused by the malware's attack. The test bed is instrumented in such a way that we can measure observable manifestations of this contest between malware and bonware, especially the performance parameters of the truck{, even if the performance of the malware and bonware are not well known (e.g., if the system uses components not designed by the system designer).}

The remainder of the paper is organized as follows. In {Section~\ref{sec:prior-work}}, we briefly describe prior work related to quantification of cyber resilience. In {Sections~\ref{sec:theory} and~\ref{sec:mathmodeling}}, we propose a class of parsimonious models in which effects of both malware and bonware are approximated as deterministic, continuous differentiable variables, and we explore several variations of such models. In addition, we discuss how parameters of such models can be obtained from experimental data and whether these parameters might be considered quantitative characteristics (i.e., measurements) of the bonware's cyber resilience. In {Section~\ref{sec:components}}, we introduce the experimental approach we used to obtain resiliency-relevant data; we describe various components of the overall experimental apparatus and the process of performing experiments. {Then, in Section~\ref{sec:experimental app and disc},} we illustrate the experimentation and analysis using a case study, discuss the experimental results. {Finally, in Section~\ref{sec:conclusions}, we} offer conclusions. 
\section{Prior work}\label{sec:prior-work}
A growing body of literature explores quantification of resilience in general and cyber resilience in particular.  {Resilience has been studied for decades in multiple domains and classes of systems and organisms: in sociology, biology, medicine, forestry, urban planning, economics, as well as technical artifacts. That research looked at definitions of resilience and it's differentiation from other properties of systems, manifestations of resilience, factors that enhance or degrade resilience, and means to characterize resilence, e.g., 
\cite{dillon1947resilience,hoffman1948generalized,herrman2011resilience,wu2013understanding,folke2010resilience}. 
Research interest in the field cyber resilience emerged more recently. Multiple works explored the definitions of cyber resilience, relations between cyber resilience and cyber security, proposals for improving cyber resilience (or cyber security with which it is often conflated), and methods to characterize, assess and quantify cyber resilience. Examples include\cite{dupont2019cyber,hausken2020cyber,herrington2013future,huang2022reinforcement,jensen2015challenges,zou2021cyber}  It is specifically the last topic --quantification of cyber resilience -- that is the scope of our paper, and therefore we now found on the body of prior work on quantification of cyber resilience.}  Approximately, the literature can be divided into two categories: (1) qualitative assessments of a system (actually existing or its design) by subject matter experts (SMEs) \cite{alexeev2017constructing, henshel} and (2) quantitative measurements based on empirical or experimental observations of how a system (or its high-fidelity model) responds to a cyber compromise \cite{kott2019cyber, ligo2021how}.  In the first category, a well-cited example is the approach called the cyber-resilience matrix \cite{linkov2013resilience}. In this approach, a system is considered as spanning four domains: (1) physical (i.e., the physical resources of the system, and the design, capabilities, features and characteristics of those resources); (2) informational (i.e., the system's availability, storage, and use of information); (3) cognitive (i.e., the ways in which informational and physical resources are used to comprehend the situation and make pertinent decisions); and (4) social (i.e., structure, relations, and communications of social nature within and around the system). For each of these domains of the system, SMEs are asked to assess, and to express in metrics, the extent to which the system exhibits the ability to (1) plan and prepare for an adverse cyber incident; (2) absorb the impact of the adverse cyber incident; (3) recover from the effects of the adverse cyber incident; and (4) adapt to the ramifications of the adverse cyber incident. In this way, the approach defines a 4-by-4 matrix that serves as a framework for structured assessments by SMEs.     

Another example within the same category (i.e., qualitative assessments of a system by SMEs) is a recent, elaborate approach proposed by \cite{beling2021developmental}. The approach is called Framework for Operational Resilience in Engineering and System Test (FOREST), and a key methodology within FOREST is called Testable Resilience Efficacy Elements (TREE). For a given system or subsystem, the methodology requires SMEs to assess, among others, how well the resilience solution is able to (1) sense or discover a successful cyber-attack; (2) identify the part of the system that has been successfully attacked; (3) reconfigure the system in order to mitigate and contain the consequences of the attack. Assessment may include tests of the system, although the methodology does not prescribe the tests. 

Undoubtedly, such methodologies can be valuable in finding opportunities in improvements of cyber-resilience in a system that is either at the design stage or is already constructed. Still, these are essentially qualitative assessments, not quantitative measurements derived from an experiment.  {As such, these are not the lines of inquiry we pursue in this paper. } 

In the second category (i.e., quantitative measurements based on empirical or experimental observations of how a system, or its high-fidelity model, responds to a cyber compromise), most approaches tend to revolve around a common idea we call here the area under the curve (AUC) method \cite{hosseini2016review, kott2021to}.

\begin{figure}[ht]
    \centering
    \includegraphics[scale=1.2]{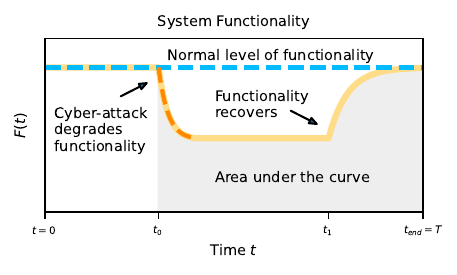}
    \caption{The functionality $F(t)$ of a system is plotted in yellow and normal functionality is plotted in light blue.  In this example, there is a cyber attack at time $t_0$ which degrades functionality, and the system's resilience allows the functionality to begin recovering at time $t_1$.  The area under the curve (AUC) is defined as the normalized area under the curve from time $t_0$  to time $T$, the end time of measured system performance.}
    \Description{The functionality $F(t)$ of a system is plotted in yellow and normal functionality is plotted in light blue.  In this example, there is a cyber attack at time $t_0$ which degrades functionality, and the system's resilience allows the functionality to begin recovering at time $t_1$.  The area under the curve (AUC) is defined as the normalized area under the curve from time $t_0$  to time $T$, the end time of measured system performance.}
    \label{fig:tub}
\end{figure}

The general idea is depicted in Fig.~\ref{fig:tub}. The functionality is plotted over time $t$. At time $t = t_0,$ a cyber attack begins to degrade the functionality of the system, as compared to the normal level of functionality. The system resists the effects of the cyber attack, and eventually stabilizes the functionality at a reduced level. At $t = t_1,$ the system resilience mechanisms begin to overcome the effect of the attack and eventually recover the functionality to a normal level. The area under the curve (AUC) reflects the degree of resilience -- the closer AUC is to its normal level, the higher is the system's resilience.

In an experiment/test, a system engages in the performance of a representative goal, and then is subjected to an ensemble or sequence of representative cyber attacks. A goal-relevant quantitative functionality of the system is observed and recorded. The resulting average functionality, divided by normal functionality, can be used as a measure of resilience.  {In this paper, we build on and extend the idea of AUC, and aim to evaluate its feasibility and efficacy for quantifying resilience. In particular, we are not aware of prior work where the idea of AUC has been experimentally investigated in the context of cyber resilience. }  

However, AUC-based resilience measures are inherently cumulative, aggregate measures, and do not tell us much about the underlying processes. For example, is it possible to quantify the resilience \impact{} of the bonware of the given system? Similarly, is it possible to quantify the \impact{} of malware? In addition, is it possible to gain insights into how these values of \impactfulness{} vary over time during an incident? We will offer steps toward answering such questions in addition to evaluating the AUC as a resilience measure.

With respect to experimental approaches, much of the early experimental work on the cybersecurity of automobiles used actual vehicles \cite{hoppe2007sniffing, koscher2010experimental, miller2013adventures, miller2015remote, foster2015fast}. This approach offers high fidelity but also high costs, especially when multiple experimental runs are required.

Other approaches avoided the expensive use of actual vehicles by connecting multiple electronic control units (ECUs) together on a Controller Area Network (CAN) bus independent of a vehicle \cite{ daily2016towards, bozdal2018hardware, wang2018delay}.  This is an inexpensive method to test malware and bonware in a vehicular network; however, it cannot characterize impacts on the vehicle's performance parameters.

Yet another experimental approach is to use a Digital Twin: a system to reproduce real-world events in a digital environment, e.g., \cite{shikata2019digital}. A virtualized vehicle with realistic virtual performance would provide high fidelity at low cost in terms of time to test and measure cyber resilience. However, constructing a virtual vehicle can be prohibitively expensive, too.  {In our research, we strove to find an effective compromise that minimizes disadvantages on the above approaches. We use real, physical electronic and computing elements of a vehicle (thereby improving the fidelity of cyber effects) and integrate it with virtual model (a digital twin) of the remainder of the vehicle. This physical-digital twin of a vehicle combines fidelity and relative affordability. }
To summarize, our work differs from prior work in the following aspects. Unlike much prior work, our focus is on quantitative, not qualitative characterization of resilience. Unlike the prior work that aims at quantitative characterization, we propose specific parsimonious mathematical models to aid in quantifying systems’ resilience to cyber attacks. Furthermore, building on the models, we propose and implement an experimental method and test bed for obtaining resilience-relevant data of an example system in repeatable, systematic experiments. Unlike prior work, we identify quantitative characteristics obtainable from experimental data, and show that these characteristics can serve as useful quantitative measures of cyber resilience. With respect to practical contributions to the field, we have proposed a new practical method for measuring cyber resilience of a system under test. The details of the method go beyond the scope of this paper, but are described in a separate publication \cite{kottmethodology}. 
{In what follows, we will first describe our formal treatment of cyber resilience, with a focus on quantification of key concepts.  We will then describe how changes in performance over time may be expressed through mathematical models, which will then be used to estimate other features of the cyber-physical system, such as the effectiveness of its bonware.}
\section{Quantitative Measurement of Cyber Resilience}\label{sec:theory}

In this section, we will first formalize our thinking about cyber resilience, and then use our new formalism to define the AUC-based measures of resilience as well as the mathematical models that we will apply to our experimental runs.  Table~\ref{tab:terms} lists variables used in the discussion along with their defintions.

\begin{table}[h]
  \caption{A Short Table of Terms and Definitions}
\begin{center}
\begin{tabular}{|l| l |} 
 \hline
 Term & Definition \\ [0.5ex] 
  \hline\hline
  $\ma$ & Accomplishment:  Cummulative sum of functionality over time\\ \hline
  $\ma_\text{attack} $ & Accomplishment achieved during the course
                           of an attack\\ \hline
  $\ma_\text{baseline} $ & Baseline accomplishment achieved when no
                             attack is present\\ \hline
  $\text{AUC}$ & Area under the curve:  A normalized measure of accomplishment\\ \hline
  $\bonware$ & Bonware effectiveness\\ \hline
  $\functionality $ & Functionality of system being studied\\ \hline
  $F_\text{attack} $& Functionality while system is under attack\\ \hline
  $F_\text{baseline} $ & Baseline functionality\\ \hline
  $\frac{dF_\bonware}{dt} ^{\phantom{0}}_{\phantom{0}}$ & Time rate of
  change of functionality due to bonware present in the system\\ \hline
  $\frac{dF_\malware}{dt} ^{\phantom{0}}_{\phantom{0}} $ & Time rate
                                                           of change
                                                           of
                                                           functionality
                                                           due to malware attacking the system\\ \hline
  $\Fnominal $ & Nominal functionality \\ \hline
  $\malware$ & Malware effectiveness\\ \hline
  $\Qware $ & The sum of malware effectiveness and bonware effectiveness\\ \hline
  $R$ & A measure of resilience computed as the ratio of
        accomplishment under attack to baseline accomplishment\\ \hline
\end{tabular}
\end{center}
\label{tab:terms}
\end{table}

\subsection{Formal Definition of Concepts}
We define goal-relevant resilience as the ability of a system to accomplish its goal---or at least maximize the degree of accomplishment of its goal---in spite of effects of a cyber attack, as a run unfolds over time. To this end, we postulate that for a given run, there exists a function $\ma(t)$ that represents accomplishment and that is cumulative from the run start time $t_0$ up until the present time $t$.  We define functionality, $\functionality(t)$, to be the time derivative of goal accomplishment.  Thus,
\begin{equation}
  \label{eq:ma}
  \functionality(t) = \frac{d \ma }{dt}, \quad \ma(t) = \int_{t_0}^t \functionality(\tau) \, d\tau.
\end{equation}
Note that, in practice, functionality may vary with time, even when the system performs normally and is not experiencing the effects of a cyber attack.  To be able to account for this, we will often distinguish between performance under baseline conditions, $F_\text{baseline}(t)$ and performance during an attack scenario, $F_\text{attack}(t)$. We will require $F_\text{baseline}(t)>0$ everywhere where it is defined.

{The purpose of this definition is to take a step towards a more formal treatment of the term `functionality', unlike in prior work where the term 
is introduced informally and is merely illustrated by domain-specific examples.}  {Also note that we use the expression ``goal-relevant" resilience in order to emphasize that the same system may be assigned different goals (e.g., a truck might be sent to deliver maximum load at reduced speed or alternatively to deliver a modest load in shortest time), and resilience does depend on a goal the system is pursuing. Needless to say, cyber resilience of a system does depend on cyber attacks imposed on the system. That is, a system may have excellent resilience against one attack, and poor resilience against another attack. Even with the same attack and the same goal pursued under different conditions, the system may exhibit different level of resilience (see Fig.~7 where cyber resilience of a truck depends on roads and terrain). Therefore, for practical applications, cyber resilience should be measured in the context of defined ensembles of attacks and scenarios of use (see \cite{kottmethodology} for discussion of practical formulation of such ensembles). Furthermore, a goal may imply multiple objectives as we discuss in }{Subsection 3.2.} {At the same time, the general approach and methodology for measuring resilience does not appear to require changes for applications to different types of systems and scenarios.}
\subsection{Resilience Based on Area Under the Curve}\label{sec:auc based}

{As mentioned in Section~\ref{sec:prior-work}, our research is based on -- and aims to extend and experimentally investigate -- the idea of area under the curve (AUC) that has been frequently discussed in prior work. Note that when considered over the entire period of a system performance (from $t_0$ to $T$), AUC is precisely}
\begin{equation*}
  \text{AUC} = \frac{1}{T-t_0} \int_{t_0}^T F(\tau)\, d\tau.
\end{equation*}

Here we expand on this concept to make a measure of resilience that calculates the accomplishment---that is, the area under the functionality curve---in a cyber attack scenario relative to the accomplishment in a baseline scenario:
\begin{equation}
\label{eq:R}
R = \frac{\int_{t_0}^{T} \functionality_\text{attack}(\tau) \, d\tau}{\int_{t_0}^{T} \functionality_\text{baseline}(\tau) \, d\tau} = \frac{\ma_\text{attack}(T)}{\ma_\text{baseline}(T)}.
\end{equation}
{As a measure of resilience, $R$ has a number of advantages.  By contrasting behavior in an attack scenario to behavior in a comparable baseline scenario, it is able to account for idiosyncratic differences between vehicles, terrain, or any other features we hold constant between the two scenarios.  To illustrate, consider Figs.~6 and~7 later in the article, where we use the fuel efficiency of trucks as the performance measure influenced by cyber attacks. Imagine two trucks -- one twice bigger and heavier than another, but otherwise with exactly the same design of cyber-relevant systems. The lighter truck would have approximately twice greater fuel efficiency (in km per liter) and the impact of a cyber attack on its AUC would be twice greater. This might imply that the cyber resilience of the lighter truck is twice worse, which is incorrect. On the other hand, the dimensionless $R$ would be roughly the same for both trucks, as it should be. $R$ can be interpreted as the \textit{fraction of normal functionality maintained} during a cyber attack.  If it is close to $1.0$, then the effect of the attack was small; if it is $0.0$, then functionality was completely disrupted. However, it is important to note that $R$ is closely related to AUC, as it reflects the ratio of AUC occurring in presence of a cyber attack to the AUC without an attack. In other words, $R$ is a normalized and dimensionless variation of the AUC measure.}

\section{Mathematical Modeling}\label{sec:mathmodeling}

Here we introduce a class of parsimonious models in which effects of both malware and bonware on goal accomplishments are approximated as deterministic, continuous differentiable variables. Our models describe the behavior of a system's functionality over the course of a run during which it is being attacked by malware and defended by bonware.
To simplify our modeling, we assume the normal functionality to be constant in time, $\Fnominal(t)=\Fnominal=1$.  Normal functionality here refers to the value of a system characteristic exhibited under baseline conditions without the impact of a cyber attack. For example, in Fig.     \ref{fig:model fits}, the cyan line describes fuel efficiency of a truck measured in a baseline run without a cyber attack. Naturally, it depends on time-varying conditions (e.g., the terrain profile, turning and acceleration of the truck) and is not constant, in general.\color{black}  When we apply our mathematical models to our experimental results in Section~\ref{sec:experimental app and disc}, we will ensure this assumption by explicitly dividing the functionality during a cyber attack scenario, $F_\text{attack}(t)$, by the functionality during a baseline scenario, $F_\text{baseline}(t)$, to obtain $F(t)$.  {Without loss of generality, for ease of notation, in Subsections 4.1, 4.3, 4.5, and 4.7 we assume $t_0=0.$}

In the first set of models, we assume that there is an observable, sufficiently smooth function representing goal accomplishment, and we define functionality to be its time derivative.  Then, we motivate a parsimonious model for the differential equation governing functionality, give the general solution, and discuss a few specific cases.

\subsection{Linear Differential Equation and General Solution}\label{sec:continuous}

{In what follows we develop a differential equation to model system functionality and how it is influenced by malware and bonware.  Imposing the least restrictions on functionality, we require functionality to be continuosly differentiable $(F \in C^1)$, and thus, accomplishment, obtained by integrating functionality over an interval of time, must be twice continuously differentiable 
$(\ma \in C^2)$.  Although this does not allow jumps or corners or the like in accomplishment, the restriction to twice continuously differentiable functions is reasonable for a continuous differential equation model.}  
{We do not anticipate that this assumption will exclude any practical systems, because even in systems where, e.g., functionality might have near instantenous changes due to a cyber attack, it can be approximated by a continuously differentiable function sufficiently closely for the purposes of our research. Indeed, we did not observe any difficulties of such nature in our experimental data.}
{When considering more sophisticated models, such as stochastic differential equations with jump processes, we may relax this assumption.}

{Both malware \impact{} and bonware \impact{} are continuous functions of time: $\malware, \bonware \in C^0.$}  
{We further assume that malware degrades the system's functionality} while bonware aims to increase functionality over time.  {Thus, both $\malware$ and $\bonware$ are nonnegative.}  {Later, in Subsections 4.3 and 4.5, we will model $\malware$ and $\bonware$ as piecewise continous functions and allow them to jump at a finite set of points in time.}  {Although it is possible that a malware inadvertently increases system's functionality (perhaps due to an error in the  malware's design) we expect such cases to be very unusual and do not consider them in this paper. The same applies to an errant bonware that happens to decrease the system's functionality.}  {We model the effectivenesses of malware and bonware independently, and relate them by adding their effects to give us the overall change in functionality due to these two competing quantities.}  We define the \impact{} of malware, $\malware{},$ to be a function that,
when multiplied by the functionality at the present time, causes the time rate of change in functionality to decrease by that amount:
\begin{equation}
  \frac{dF_\malware(t)}{dt} = - \malware(t) F(t).
\end{equation}
{Similarly, we define the effectiveness of bonware $\bonware$ via the following equation.} 
\begin{equation}
  \frac{dF_\bonware(t)}{dt} = \bonware(t) (\Fnominal (t) - F(t)).
\end{equation}
{With this definition, $\bonware$ serves to restore the functionality}  by causing the time rate of change in functionality to increase by the product of $\bonware$ with the difference between normal and current functionality.  The \impact{} on functionality is the sum of the \impacts{} of malware and bonware:  \begin{equation*}\frac{dF(t)}{dt}= \frac{dF_\malware(t)}{dt}+\frac{dF_\bonware(t)}{dt}. \end{equation*} Thus
\begin{equation}
    \frac{d\functionality}{dt} + \Qware(t) \functionality(t) = \Fnominal \bonware (t), \quad \text{ where }\Qware(t)=\malware(t)+\bonware(t).
    \label{eq:00}
\end{equation}

Since we expect bonware to help (or at least not harm) and malware to not help, we assume $\bonware(t) \ge 0$ and $\malware(t) \ge 0$.  We also assume normal functionality is positive, $\Fnominal > 0,$ 
{i.e., that the system was designed in its normal operation to increase the accomplishment of the system's intended goals over time (note Eq.~(1))}
and functionality is always positive and less than or equal to normal functionality, $0 < \functionality(t) \le \Fnominal.$ 
{Although it might be possible that a cyber attack subverts a system to such an extent that its functionally becomes negative, i.e., it works against its intended goals, we exclude such cases from the scope of this paper.} 

{Under the assumptions, the first order linear differential equation Eq.~(5)}
 
has the following solution: 

\small
\begin{equation*}
    F(t)  = \expminus \left( F(0) + \bintegral \right).
\end{equation*}
\normalsize
In Appendix \ref{appendix:A}, we will discuss the stability of Eq.~(\ref{eq:00}) and show that it is stable due to the non-negativity of both $\malware(t)$ and $\bonware(t)$.  est{We have not yet explored the possibility of an adversary intruducing multiple switchings of $\malware$.  It was suggested to us, by our reviewers that this could cause an instability.  Proposition 2 in \cite{lin} may prove helpful to those wishing to replicate results here and/or study more sophisticated attacks.}
To help us understand how the model works, {in Sections 4.2---4.5} we find explicit solutions for a number of examples.

\subsection{Constant model}

Assuming $\malware, \bonware,$ and $\Qware$ are constant, we have
\begin{equation}  \label{eq:1}
    \frac{d\functionality}{dt} + \Qware \functionality(t) =  \Fnominal \bonware.
\end{equation}

\subsubsection{No bonware}

If $\bonware=0$, then Eq.~(\ref{eq:1}) reduces to $\frac{d\functionality}{dt} + \malware \functionality(t) =  0$ and $\functionality(t) = \functionality(0) e^{-\malware t}$. If also $\malware=0$ (no bonware and no malware), then $\frac{d\functionality}{dt}=0$ and $\functionality(t)=\functionality(0)$.

\subsubsection{Bonware}\label{sec:4.2}

With bonware present, the solution is 
\begin{equation} \label{eq:3}
    \functionality(t) = \left(\functionality(0) - \frac{\Fnominal \bonware}{ \Qware } \right) e^{-\Qware t} + \frac{\Fnominal \bonware}{\Qware}.
\end{equation}

\begin{figure}[ht]
    \centering
    \includegraphics[scale=1.2,trim=5 5 5 7,clip]{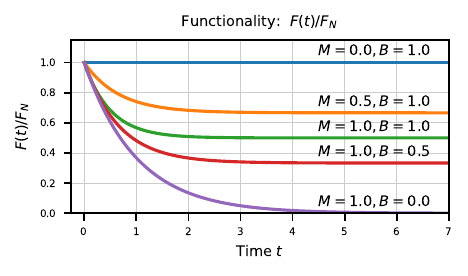}
    \caption{Normalized functionality, $\functionality(t)/\Fnominal$, is shown for various constant values of $\malware$ (malware attacking) and $\bonware$ (bonware defending) and with initial condition $\functionality(0)=\Fnominal$.  The functionality over time depends on the relative strengths of bonware and malware.  With the system initially at normal functionality and malware \impact{} nonzero, functionality exhibits exponential decay.}
    \Description{Normalized functionality is shown for various values of malware effectiveness (malware attacking) and bonware effectiveness (bonware defending) and initial condition.  The functionality over time depends on the relative strengths of bonware and malware.  With the system initially at normal functionality and malware effectiveness nonzero, functionality exhibits exponential decay.}
    \label{fig:2}
\end{figure}

If $\functionality(0)> \large \sfrac{\Fnominal\bonware}{\Qware},$ then $\functionality(t)$, initially at $\functionality(0)$ at time $t~=~0$, will decrease to $\large \sfrac{\Fnominal\bonware}{\Qware}$ (see Fig.~\ref{fig:2}).  If  $\functionality(0)~=~\large \sfrac{\Fnominal\bonware}{\Qware},$ then the function $\functionality(t) = \functionality(0)$ will be constant.  If $\functionality(0)<\large \sfrac{\Fnominal \bonware}{\Qware},$ the function will start at $\functionality=\functionality(0)$ and increase to $\large \sfrac{\Fnominal \bonware}{\Qware}.$ The plots for $\malware>0$ in Fig.~\ref{fig:2} show that even in the presence of bonware, malware has an impact on the system.  \blue{The steady-state of the system is obtained either by setting $\frac{d\functionality}{dt}=0$ in Eq.~(\ref{eq:1}) or letting $t\to\infty:$}
\begin{equation} \label{eq:5}
    \functionality_\infty=  \lim_{t\to\infty} \functionality(t)  =\Fnominal\frac{\bonware}{\malware+\bonware}
\end{equation}
so that the antidote to malware is to overwhelm it with bonware.  \blue{The exponent, $-\Qware t = (-\malware-\bonware)t$ in the solution given by Eq.~(\ref{eq:3}) indicates that increasing the \impact{} of either malware or bonware will cause the system to more quickly approach steady-state.}{}  At steady-state,
\begin{equation}\label{eq:5a}
    \begin{aligned}
       {\frac{ \Fnominal -\functionality_\infty}{\Fnominal}  =  \frac{\malware}{\malware+\bonware}. }
    \end{aligned}
\end{equation}
Eq.~(\ref{eq:5a}) gives us further insight into the trade-off between \impacts{} of both malware and bonware.  The relative decrease of the function from normal functionality is equal to the ratio of malware \impact{} to the sum of malware and bonware \impacts{}.
\subsection{Piecewise constant model}\label{subsection:piecewise constant model}

If either malware's or bonware's \impact{} diminishes at some point in the incident, the model may switch from one set of constants defining malware and bonware to another set of constants.  The differential equation (Eq.~(\ref{eq:00})) may now be expressed as
{
\begin{equation}
    \frac{d\functionality}{dt} = \sum_{j=0}^{N-1}\left[ (\Fnominal-\functionality(t)) \bonware_j -  \functionality(t) \malware_j
    \right]\Pi_{t_j t_{j+1}}(t),
    \label{eq:000}
\end{equation}
}
where the vectors $\boldsymbol{\malware} =( {\malware}_0, {\malware}_1, \hdots, {\malware}_{N-1} )$ and $\boldsymbol{\bonware} =( {\bonware}_0,  {\bonware}_1,\hdots, {\bonware}_{N-1})$ contain the malware \impacts{} and bonware \impacts{} within time windows whose end points are defined by  $\{t_0, t_1, \hdots, t_N \}$ and where 
{
\begin{equation*}
\Pi_{ab}(t) = \left\{ 
\begin{aligned}
1, & \quad a \le t < b, \\
0, & \quad \text{otherwise.}
\end{aligned}
\right.
\end{equation*} 
}
{The assumption $\functionality \in \mathbf{C}^1$ holds within each time interval.  Functionality $\functionality(t)$ is continuous, but we no longer assume $\functionality(t)$ is differentiable at the time window endpoints $\{t_0, t_1, \hdots, t_N \}$. }
The solution will be a function which, in each time interval, is the solution found in Eq.~(\ref{eq:3}):
\begin{equation*} \label{eq:piecewise:constant}
    \functionality(t)     = \left(\functionality(t_j) - \frac{\Fnominal \bonware_j}{ \Qware_j } 
    \right) e^{-\Qware_j (t-t_j)} + \frac{\Fnominal \bonware_j}{\Qware_j},                 
    \quad (t_j \le t < t_{j+1}), \quad (j=0,\hdots, N-1)                                        
\end{equation*}
where $\Qware_j = \malware_j + \bonware_j$.
{The model is general enough so that we have the ability to handle instantiations where malware's and bonware's effectiveness are out of sync.  For example, If $t_1 <t_2<t_3<t_4$ and malware's effectiveness is nonzero in time window $(t_1, t_3)$ and bonware's effectiveness is nonzero in time window $(t_2,t_4)$, then we have}
   \begin{equation*}
      {\malware} = \left\{
          \begin{aligned}
            \malware_1 (t), & \quad t_1 \le t < t_2\\
            {\malware}_2 (t), & \quad t_2 \le t< t_3\\
              0, & \quad t_3 \le t <  t_4.
          \end{aligned}
          \right.
          \quad
          \text{ and }
          \quad
      {\bonware} = \left\{
          \begin{aligned}
            0, & \quad t_1 \le t < t_2\\
            {\bonware}_2 (t), & \quad t_2 \le t< t_3\\
            { \bonware}_3(t), & \quad t_3 \le t <  t_4.
          \end{aligned}
          \right.
      \end{equation*}

{\subsection{Linear time varying (LTV) model} } \label{sec:linear}

\newcommand{\tempa}{\Omega(t)}
\newcommand{\tempb}{\Lambda}

The \impacts{} of malware and bonware may also be linear functions of $t$, so that $\malware(t) = \nu-\mu t$, $\bonware(t)~=~\alpha~-~\beta~t$, and $\Qware(t) = \lambda - \omega t$, where $\lambda = \alpha + \nu$ and $\omega  = \beta + \mu$.  Under this linear time varying model, Eq.~(\ref{eq:00}) becomes
\begin{equation}
    \frac{d\functionality}{dt} + (\lambda - \omega t) \functionality(t) = \Fnominal (\alpha - \beta t).
    \label{eq:linear:model}
\end{equation}
The solution is derived in Appendix \ref{appendix:B} and is expressed in terms of the error function
$\blue{\erf(z)=\frac{2}{\sqrt{\pi }}\int_0^z e^{-\tau^2}\,d\tau}$:
\begin{equation}
    \label{eq:6}
        \frac{\functionality(t)}{\Fnominal} 
            = \frac{1}{\tempa} \left\{                                
                \frac{\functionality(0)}{\Fnominal}
               -\frac{\beta}{\omega}\left(1-{\tempa}\right) +(\alpha \omega
                -\beta \lambda )
            \right.\\
            \left.  \frac{\sqrt{\frac{\pi }{2}} e^{\tempb^2} }{\omega^{3/2}}\left[\erf\left(\tempb\right)+\erf\left(\frac{\omega t}{\sqrt{2 \omega }}-\tempb\right)\right] \right\}
\end{equation}
\normalsize
where $\tempa = e^{\lambda  t-\frac{1}{2}\omega t^2}$, and $\tempb = \large \sfrac{\lambda}{\sqrt{2 \omega }}$.

The LTV model is helpful when the effectiveness of malware $\malware$ and of the bonware $\bonware$ are dependent.  The dependency may or may not be strong. In our experiments, we did not notice such a dependency, but in other systems it might be too strong to ignore it.  Although the LTV model does not capture such a dependency explicitly, it may help to reflect the results of the dependency.  Future research should address the ways to model the dependency and to incorporate it into our approach.

\subsection{Piecewise linear time varying (PLTV) model}

\newcommand{\tcj}{\Omega_j(t)}  
\newcommand{\td}{\Lambda}

Both malware and bonware \impacts{} may initially be linear, but if the situation changes and a different linear model holds after a time, the model should be able to account for it.  In particular, if malware \impact{} is decreasing over time, at some point we will reach $\malware=0$ and the model switches to a new linear model. Eq.~(\ref{eq:linear:model}) can be written
{
\begin{equation*}
    \frac{d\functionality}{dt} = \sum_{j=0}^{N-1}  \left[ (\lambda_j - \omega_j t) \functionality(t) -  \Fnominal (\alpha_j - \beta_j t) \right]\Pi_{t_j t_{j+1}}(t). \label{eq:piecewise:linear}
\end{equation*}
}
The solution follows from Eq.~(\ref{eq:6}):
\small
\begin{equation*} \label{eq:piecewise:linear:solution}
    \frac{\functionality(t)}{\Fnominal} 
    = \frac{1}{\tcj} 
    \left\{ 
    \frac{\functionality(t_j)}{\Fnominal}-\frac{\beta_j}{\omega_j }\left(1-\tcj\right) +(\alpha_j  \omega_j -\beta_j  \lambda_j )\right.\\
    \left. \frac{\sqrt{\frac{\pi }{2}} e^{\td_j^2 }}{\omega_j^{3/2}} \left[\text{erf}\left(\td_j\right)+\text{erf}\left(\frac{\omega_j (t-t_j) }{\sqrt{2 \omega_j }}-\td_j\right)\right] \right\}, \\
                                        \quad \!\!\! (t_j \le t < t_{j+1}), \!\! \quad (j=0,\hdots, N-1)
\end{equation*} \normalsize
where $\tcj=e^{\lambda_j (t-t_j)-\frac{1}{2}\omega_j (t-t_j)^2 }$ and $\td_j=\large \sfrac{\lambda_j }{\sqrt{2 \omega_j }}$.  

\begin{figure}[ht]
    \centering
    \includegraphics[scale=1.2]{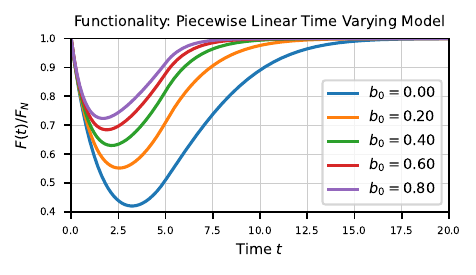}
    \caption{Normalized functionality, $\functionality(t)/\Fnominal$,
        is shown for piecewise linear time varying models and initial condition
        $F(0)=\Fnominal$. Both malware and bonware \impacts{} are
        initially linear functions of time: $\malware~=~\max(0.5-0.1t,0), \bonware = b_0+0.04t.$ When malware \impact{} reaches $\malware=0$, bonware \impact{} continues to increase.}
        \Description{Normalized functionality 
        is shown for piecewise linear time varying models and initial condition. Both malware and bonware effectiveness are
        initially linear functions of time.  When malware effectiveness reaches zero bonware effectiveness continues to increase.}
    \label{fig:piecewise:linear}
\end{figure}

Example realizations of the piecewise linear time varying models are shown in Fig.~\ref{fig:piecewise:linear}.

\section{Experimental Testbed and Method}\label{sec:components}

A key role of the mathematical models presented above is to help analyze results of actual experimental measurements of resilience. In this section, we introduce the experimental test bed and experimental process we use to observe and characterize cyber resilience of a truck \cite{ellis2022experimental}. In a typical resilience-measuring experiment, the following occurs, conceptually: (1) The truck is assigned a goal (delivering a cargo to a destination, along a specified route). The truck begins to accomplish the goal. The driver controls the truck aiming to maximize probability of the goal's success. (2) At some point along the route, an adversarial cyber effect is activated and begins to degrade goal-relevant performance of the truck. (3) Physical and cyber elements within the truck begin to resist the impact of the cyber effect. After some time, these elements (i.e., bonware, collectively) may succeed in recovering some or all of the degraded performance. (4) The data collection and logging system obtains and records the performance parameters of the truck over time, from the beginning of the run until its end (successful or otherwise). The data are later analyzed using the models presented earlier. 

\begin{figure}[ht]
    \includegraphics[width=.75\columnwidth,trim=0 15 0 10,clip]{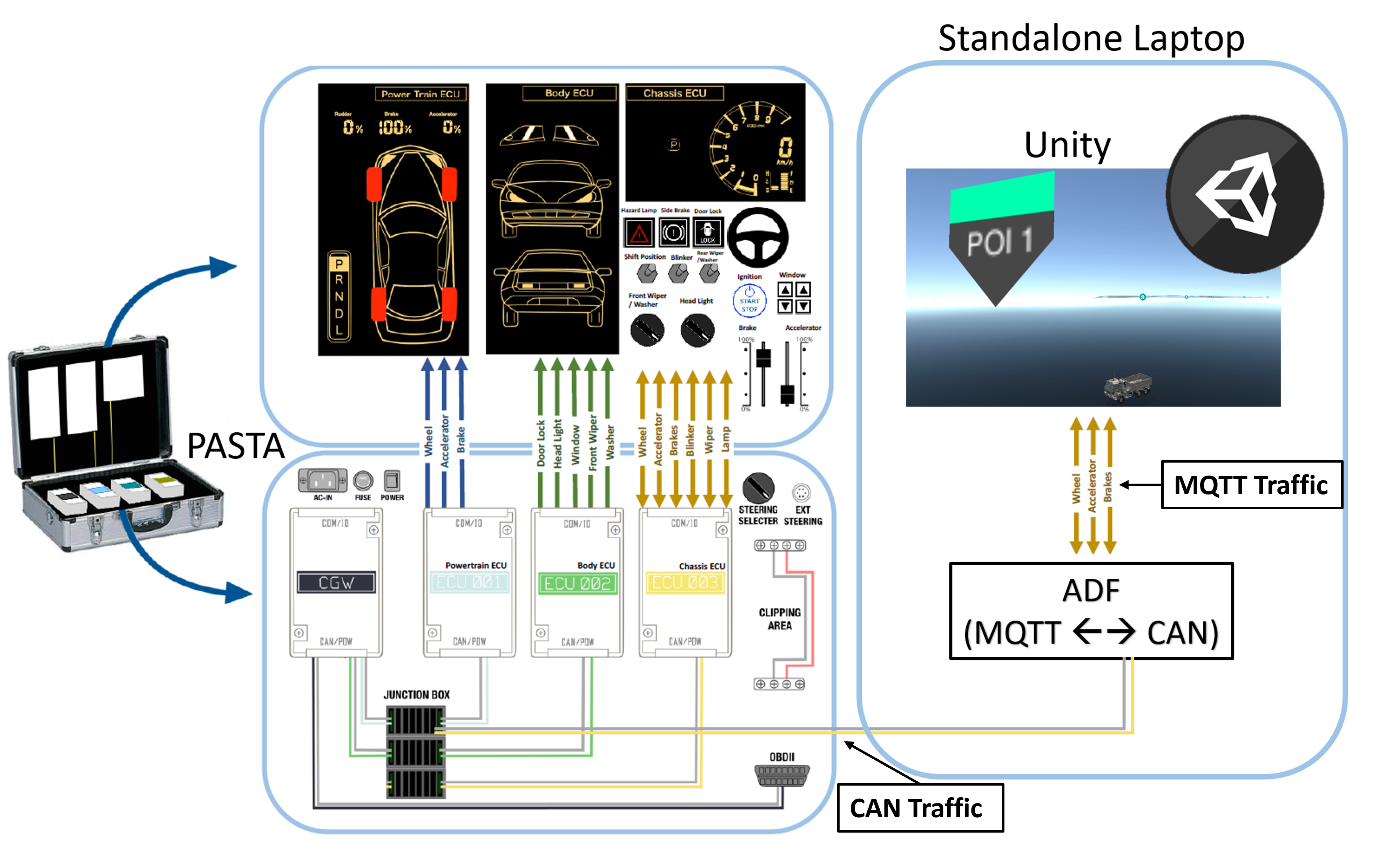}
    \caption{
{
  A high-level overview of the data flow between components. Portions are derived from \cite{toyama2018pasta}. Unity handles the interactions between the simulated trucks and the terrain, provides driver inputs to maintain pathing and speed over a pre-determined course across the terrain, and simulates many aspects of the vehicle to provide high-fidelity vehicle performance. PASTA is hardware-in-the-loop that generates actual CAN bus traffic in response to the inputs from Unity, and provides three vehicle ECUs (powertrain, body, and chassis). ADF allows for communication translation between Unity and PASTA, provides simulated ECU plugins for additional sensor information not present in PASTA, and implements simulated cyber attacks and defenses.}
}
    \Description{
        A high-level overview of the data flow between components. Portions are derived from Toyama's paper.
    }
    \label{fg:datamodel}
\end{figure}

These processes and functions are implemented in several components of the test bed, which include automotive hardware and simulation software: the \ac{PASTA} by Toyota Motor Corporation, the Unity game development platform, \ac{ADF} developed at the DEVCOM Army Research Laboratory, and the OpenTAP test automation framework by Keysight Technologies. These components and their roles are described in Subsections~5.1---5.4 below. In terms of interactions between the components, Unity generates messages via the \ac{MQTT} publish-subscribe network protocol. \ac{ADF} ingests these messages and translates them to \ac{CAN} format, which are then injected onto the appropriate \ac{CAN} bus within \ac{PASTA}. Fig.~\ref{fg:datamodel} illustrates the flow of data between components.  

\subsection{PASTA}\label{subsection:pasta}

\ac{PASTA} is a cyber-physical product by Toyota, intended to develop and evaluate new vehicle security technology and approaches on realistic ``white-box'' \acp{ECU} \cite{toyama2018pasta}. There are three vehicle \acp{ECU} provided within the product, each with its own \ac{CAN} bus: powertrain, body, and chassis. These three \acp{ECU} are responsible for their respective group of messages, each generating and responding to traffic on their bus. A fourth \ac{ECU}, the \ac{CGW}, acts as a junction between the three previously mentioned buses. Based on the message and source bus, the \ac{CGW} ferries messages to their appropriate destination bus. The firmware for all \acp{ECU} is open-source, and is accompanied by an \ac{IDE}.

\ac{PASTA} includes simulation boards which calculate how the current \ac{CAN} messages on the buses would physically influence a commercial sedan. These boards then update the vehicle \acp{ECU} with appropriate values. For example, when acceleration pedal operation is inputted, the chassis \ac{ECU} sends a message with the indicated value. The simulation boards observe this message and calculate the physical effects that would result from the input. The results are used to update the values reported by the powertrain \ac{ECU}, which it outputs onto its bus. In this instance, the powertrain \ac{ECU} would send messages indicating the new engine throttle position, \ac{RPM}, and speed. 
Unfortunately, we found the simulation boards rather constraining, for our purposes. We cannot alter, for instance, the parameters involving the engine (e.g., torque and horsepower), the weight of the vehicle, or the terrain that the boards are assuming is being traversed. To overcome these constraints, we integrated a simulation engine (Unity, see below) that would allow for user-defined vehicle details as well as custom terrain. In this configuration, \ac{PASTA} becomes hardware-in-the-loop for the simulation engine. Cyber attacks or defenses that affect the \acp{ECU} present in \ac{PASTA} will also affect the performance of the truck within the simulation.

\subsection{Unity}\label{subsection:unity}

Unity is a widely used game development platform \cite{unity2021unity}. In particular, Unity provides built-in assets and classes regarding vehicle physics, which we leverage to model interactions between our simulated truck and custom terrain.

\subsubsection{Simulated Trucks}\label{subsection:sim-trucks}

We implemented three types of truck within Unity – light, medium, and heavy. They are designed to interface with data inputs from the white-box \acp{ECU} within \ac{PASTA}. In an experiment, the current chosen truck produces inputs in response to the simulated terrain. These inputs are sent to the corresponding \acp{ECU} within \ac{PASTA}. We then gather responses to these inputs from the \acp{ECU} and send them back to the truck, which it uses to calculate parameters like torque and fuel consumption. For example, assume the truck reports that the accelerator is set to 50\%. This message is injected into \ac{PASTA} as if the chassis \ac{ECU} had generated it. The powertrain \ac{ECU} responds to the message with the corresponding amount of engine throttle. A message with the engine throttle is sent back to Unity, which is then applied to engine power calculations. With this flow, any cyber attacks impacting the \acp{ECU} within \ac{PASTA} will affect the truck.

An automated driver is responsible for generating steering, acceleration, and braking inputs as the truck traverses the terrain. Steering is guided using a waypoint system. Both acceleration and braking inputs are calculated via a \ac{PID} controller guided by a target speed. The controller responds to changes in the terrain or truck performance, and maintains the target speed while preventing oscillation. Optional target speed variability simulates driver attention drift, which may be used to generate multiple unique realizations of the simulation under otherwise identical conditions.

Engine performance is calculated through the use of torque, horsepower, and \ac{BSFC} curves. Engine RPM is derived using the speed, wheel circumference, and effective gear ratio. Using this \ac{RPM} value, the curves are evaluated to discern the corresponding torque, horsepower, and \ac{BSFC} value. Torque is multiplied by the throttle and the current effective gear ratio to obtain the total amount of torque that can be applied to the wheels. Since our trucks are \ac{AWD}, each wheel receives the total amount of torque divided by the number of wheels on the truck. Horsepower and the \ac{BSFC} value are used in conjunction to calculate the amount of fuel that has been used per physics update.

The truck is capable of providing sensor information that is either not present in \ac{PASTA} or needs its functionality altered for our purposes. Currently, this applies to the engine coolant temperature, attitude sensor, and a set of backup engine \acp{ECU}. Engine temperature is present in \ac{PASTA}, but is aligned to the temperature characteristics of a static commercial sedan. Within Unity, we implemented a temperature model that can be controlled by an external fan controller \ac{ECU}. The fan controller monitors the coolant temperature reported by the truck and dictates the operation of a simulated fan on the truck. The fan itself takes significant power to operate, which results in a drop in the available torque that can be applied to the wheels.

\subsubsection{Terrain}

The truck within Unity traverses a custom terrain map that is roughly 81.8 km by 100 km with altitudes up to 910 m. We crafted a course approximately 151 km in length across the map that encompasses multiple terrain types: flat main road, flat off-road, hilly, prolonged ascent, and prolonged descent. On a flat main road, the target speed is 60 km per hour. Otherwise, the target speed is 40 km per hour.

\subsection{Active Defense Framework}

\ac{ADF} is a government-developed framework for prototyping active cyber-defense techniques. \ac{ADF} currently supports \ac{IP} networks and vehicle control networks, namely the \ac{CAN} bus and \ac{SAE} J1708 bus. The framework acts as an intermediary for network traffic, as depicted in Fig.~\ref{fg:datamodel}, allowing it to control network message flow, as well as inspect, modify, drop, or generate network messages. In our experimental test bed, \ac{ADF} enables communication between \ac{PASTA} and Unity by translating \ac{CAN} messages to and from \ac{MQTT}, a standard publish-subscribe \ac{IP}-based messaging protocol. \ac{ADF} plugins are also used to provide simulated \ac{ECU} hardware, and to implement cyber attack and defense methods on the CAN bus via ADF’s ability to monitor, modify, inject, or drop CAN traffic.

\subsubsection{Unity-to-PASTA Message Translation}

\ac{ADF} and Unity run on a standalone laptop and are connected to \ac{PASTA} via two \ac{USB} \ac{SLCAN} interface modules. One module is connected to the powertrain \ac{CAN} bus, and the other module is connected to the chassis \ac{CAN} bus. The \ac{PASTA} \ac{CGW} is disconnected from the CAN buses for our experiments, and the body \ac{CAN} bus and body \ac{ECU} are not used. \ac{ADF} is configured to serve as a \ac{CGW} between Unity and \ac{PASTA}. Since Unity does not natively communicate with \ac{CAN} interfaces, \ac{ADF} translates \ac{CAN} messages in real-time to \ac{MQTT} messages and back. Unlike the \ac{PASTA} \ac{CGW}, \ac{ADF} does not relay messages between the powertrain and chassis \ac{CAN} buses themselves. \ac{ADF} relays powertrain \ac{CAN} messages between Unity and \ac{PASTA}, and sends parameters from Unity to the chassis \ac{CAN} bus for display on the \ac{PASTA} instrument cluster. All communication channels are two-way.

\subsubsection{Virtual \acp{ECU} within \ac{ADF}}

{For some cyber attacks, a virtual \ac{ECU} is needed. For example, as mentioned before, the \ac{PASTA} platform does not simulate a controllable cooling fan or provide fan controller \ac{ECU} functionality. Therefore, we simulate a fan controller \ac{ECU} using an \ac{ADF} plugin. The fan controller {utilizes hysteresis control. When the engine coolant temperature is above the hysteresis band for normal operation, the controller engages the engine cooling fan. When the temperature is below, the controller disengages the fan.} For the purposes of our experiments, the fan controller \ac{ECU} plugin can simulate an attack on its own firmware, stop the attack, or reset/``re-flash'' itself (i.e., replace the \ac{ECU} firmware). During a reset, the fan controller is offline for a period of time. Use of \ac{ADF} enables creation of other simulated \acp{ECU} and corresponding cyber attacks.}

\subsubsection{Performing Cyber Attacks via \ac{ADF}}

A class of attacks on a vehicle bus involves injecting messages. Messages are broadcast on a \ac{CAN} bus, so one message injected at any point on the bus will reach all \acp{ECU} on the bus. While injection attacks cannot block or modify normal \ac{CAN} bus traffic, they can impact vehicle performance if injected messages cause undesired vehicle behavior. If an attacker can physically sever the \ac{CAN} bus wiring at a strategic point and place additional hardware there, it is possible to block or modify the normal bus traffic. Cyber attacks that block or modify messages can prevent \acp{ECU} from controlling the vehicle or falsify vehicle data.

As a man-in-the-middle between \ac{PASTA} and Unity, \ac{ADF} can execute any of these bus-level attack types.

Cyber attacks on \ac{ECU} firmware, by embedding malware, are also feasible. We have simulated the effects of embedded malware in three instances: on the fan controller, the suspension controller, and the main engine \ac{ECU}. Malware on the fan controller simulates a ``stuck fan'' attack in which malware has modified the fan control \ac{ECU} to not disengage the fan once engaged, even when the coolant temperature has dropped below the minimum operational temperature; the suspension controller attack creates the appareance that the truck is abnormally tilted, forcing the truck into a safe ``limp home'' mode that reduces the amount of available gears; the main engine \ac{ECU} attack causes erratic performance behavior.

\subsubsection{Performing Cyber Defensive Actions via \ac{ADF}}

Defending against message injection, blocking, and modification at the bus-level requires detecting and filtering injected messages before they reach the \ac{ECU}. The \ac{CAN} bus can be split at potential access points and hardware placed in-line, hardware can be placed between the \ac{CAN} bus and critical \acp{ECU}, or defenses can be integrated into the \acp{ECU} themselves. Examples of these defenses implemented previously using \ac{ADF} cryptographic watermarking and modeling observable vehicle states to compare current parameters to the model's prediction.

Cyber attacks on \acp{ECU} themselves must be approached differently. If an \ac{ECU} is compromised, measures need be taken to restore proper \ac{ECU} function. Many \acp{ECU} can be re-flashed while the vehicle is operational. The \ac{ECU} may or may not be functional for some duration while being reset or re-flashed, and the impact this will have on vehicle performance depends on the function of the \ac{ECU}. For the \acp{ECU} simulated by ADF plugins, the behavior is to make the \ac{ECU} unresponsive for a set duration, after which normal \ac{ECU} operation is restored. Note, for \acp{ECU} like the main engine \ac{ECU}, this is not possible because the vehicle will become inoperable in its absence. To address this, a manually-crafted \ac{ECU} backup is used while the main \ac{ECU} is re-flashed. {To mitigate the chance and effect of a potential supply chain attack, the backup \ac{ECU} is not a complete copy of the true main \ac{ECU} in terms hardware, software, or performance, since if it was, then the cyber attack affecting the main \ac{ECU} would also affect the backup. This means that the backup \ac{ECU} will not be as optimized as the main \ac{ECU}, thus exhibiting worse engine performance while the backup is active.}

\subsection{OpenTAP and Data Collection}

OpenTAP is an open-source test automation framework developed by Keysight Technologies \cite{opentap2021whitepaper}. It provides a test sequencer to promote test repeatability, a customizable plugin facility capable of integrating plugin classes implemented in C\# or Python, and result listeners responsible for capturing test data for further analysis. OpenTAP is used to automate the execution of experiments and provide a GUI for testing practitioners to configure experiment steps.

Data are captured from the truck. Examples of data are fuel efficiency, speed, engine torque, and acceleration pedal input; each data value comes with a timestamp of the value occurrence. 

\subsection{Execution of Experiments}

Each individual experimental run follows the same set of steps.  During setup, we establish \ac{CAN} connections to \ac{PASTA}, ensure the messaging infrastructure is running, and start Unity.  During parameter selection, we determine the truck type, cargo weight, type of cyber attack, etc., and designate the number of runs.  During execution, we run automated test scripts with the given parameters and capture the data in a desired format.  Finally, we parse and preprocess the data, fit our mathematical models, and generate graphs and results.

An experiment reflects execution of a single run as described in the beginning of this section. On our terrain, a typical run would take 2-3 hours to traverse in its entirety. However, we focus on shorter 15-minute runs that encompass a cyber attack at variable moments within the run and a potential recovery. Note that it may take the truck several minutes to recover from the attack. We are also capable of executing faster-than-real-time when using solely simulated components, further decreasing execution time of experiment runs.

\section{Experimental Approach and Discussion}\label{sec:experimental app and disc}

Using our test bed, we conducted a series of experimental runs in which a truck is subjected to a cyber attack. {To form a series of experiments, within our test bed, there are multiple parameters that can be configured at the start of a run to generate varied data captures. Currently, these include: truck type, experiment duration, cyber attack start time, terrain type(s), starting location, ending location, target speed, cyber attack-defense pairings, cargo weight, and target speed variability in the form of random seeds used in the generation of ``driver attention drift'' as described in Subsection~\ref{subsection:sim-trucks}. These random seeds are logged so that corresponding baseline and attack run pairs of the same parameter configuration use the same random seed. For example, when examining the effect of the engine \ac{ECU} attack on a light truck with medium cargo weight on a flat road, both the baseline run and the corresponding run with the attack active will use the same random seed for the target speed variability. This seed will then be be changed, and both runs re-executed to gather more corresponding baseline and attack run pairs. 
}

Here, we focus on one of these series. In this series of experiments, we considered three types of trucks with four possible cargo weights including $0$, the five unique terrains described above, and three types of cyber attacks, including one ``baseline'' scenario with no cyber attack (see Table~\ref{tab:design}).  For each combination of these, we conducted 30 experimental runs and recorded the truck's speed, fuel efficiency, and other operating parameters. The 30 runs were made unique by adding random variability to the driver's interaction with the accelerator.

\begin{table}[ht]
    \centering
    \caption{Overview of Experimental Design}
    \begin{tabular}{|p{1.2in}| p{1.7in}|}
      \hline
      Independent variable     & Possible values                                                      \\
      \hline
      \hline
        3 trucks               & \{ Light, Medium, Heavy \}                                           \\  \hline
        5 terrains               & \{ Steady Descent, Flat Road, Flat Off-Road, Hilly, Steady Ascent \} \\ \hline
        4 cyber attack scenarios & \{ Baseline, Fan, ECU, Suspension \}                                 \\  \hline
        4 cargo weights          & \{ None, Light, Medium, Heavy \}                                     \\ \hline
        30 random seeds          & \{1 \ldots 30\}                                                      \\\hline
    \end{tabular}
    \label{tab:design}
\end{table}

Considering the large scope of our experiment, we focus on examples of the types of conclusions we are able to draw in Subsections \ref{resilience results} and \ref{modeling approach}. Namely, our cyber attacks cause decreases in performance in the expected parameters at the expected times.  For example, the cyber attack on the suspension causes a reduction in fuel efficiency and speed compared to the baseline scenario. We also see that fuel efficiency decreases as cargo weight increases.

\subsection{Data Pre-processing}
\label{data preprocessing}

\begin{figure}[ht]
    \centering
    \begin{subfigure}{0.495\textwidth}
        \centering
        \includegraphics[width=\textwidth,trim=0 29 0 0,clip]{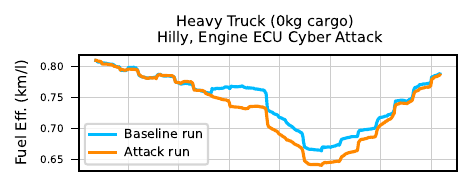}
        \includegraphics[width=\textwidth,trim=0 5 0 22,clip]{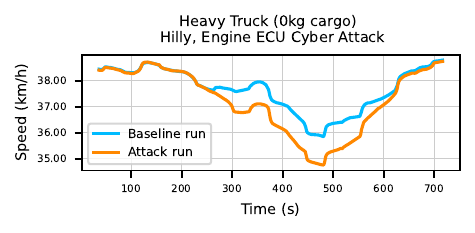}
    \end{subfigure}
    \hfill
    \begin{subfigure}{0.495\textwidth}
        \centering
        \includegraphics[width=\textwidth,trim=0 5 0 3,clip]{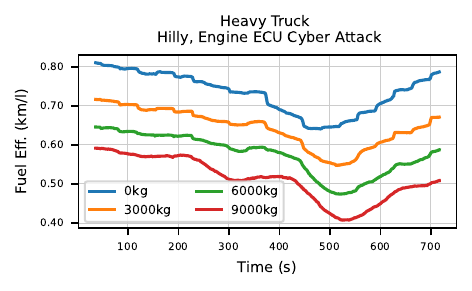}
    \end{subfigure}
    \caption{
        Examples of experimental data, illustrating that cyber attacks reduce performance both in fuel efficiency (top left panel) and speed (bottom left panel), and that changes in cargo weight reduce fuel efficiency in the expected manner (right panel). \textbf{Top left panel:} The fuel efficiency of a heavy truck, carrying no cargo, during a run on hilly terrain. The orange curve indicates the fuel efficiency in the ``engine ECU attack'' run, which is contrasted with the (partly occluded)  cyan curve that indicates the baseline run. \textbf{Bottom left panel:} Recorded speed during the same run. \textbf{Right panel:} Fuel efficiency, now for all four cargo conditions (from top to bottom: 0, 3,000, 6,000, and 9,000kg).
    }
        \Description{
        Examples of experimental data, illustrating that cyber attacks reduce performance both in fuel efficiency (top panel) and speed (middle panel), and that changes in cargo weight reduce fuel efficiency in the expected manner (bottom panel). In the top papen the fuel efficiency of a heavy truck, carrying no cargo, during a run on hilly terrain. The orange curve indicates the fuel efficiency in the ``engine ECU attack'' run, which is contrasted with the (partly occluded)  cyan curve that indicates the baseline run. In the middle panel, shown is recorded speed during the same run. In the bottom panel, fuel efficiency, now for all four cargo conditions.
    }
    \label{fig:simulated data}
\end{figure}

The operating parameters were recorded at a relatively high frequency of about 50~Hz, which sometimes causes numerical instability (e.g., in calculating fuel efficiency over a 20~ms period).  For this reason, we first applied a smoothing filter to the data.  {We chose a running median filter with a window of 72 s. This length was chosen because the engine cooling fan has a typical rhythm of turning on and off as the engine heats up with normal use. This causes the fuel efficiency to exhibit a slight sqaure-wave pattern with a period 72 s, so our chosen filter computes the median over one period of this cycle.} The running median also has the advantage that it downweights extreme values that might result from numerical inaccuracy.  We then took the mean of the 30 runs in each condition to obtain the relatively smooth time series seen in Fig.~\ref{fig:simulated data}.

The three panels of Fig.~\ref{fig:simulated data} each show the time course of a performance parameter. The left two panels each show one curve for the baseline run (cyan) and one for an attack run (orange). The right panel shows four attack runs with different cargo weights.

\subsection{Modeling Approach}
\label{modeling approach}

\begin{figure}[ht]
    \centering
   \includegraphics[width=.99\columnwidth,trim=0 5 0 22,clip]{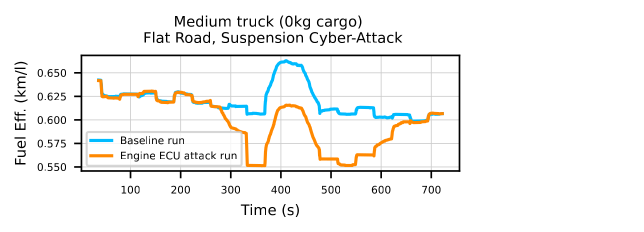}
   \includegraphics[width=.99\columnwidth,trim=0 5 0 22,clip]{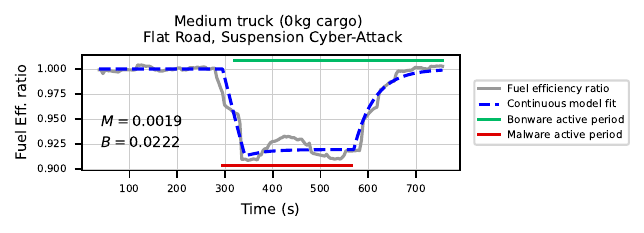}
    \caption{
        Progression of data over time. 
        \textbf{Top panel:} Fuel efficiency of a heavy truck, carrying no cargo, during a run on steadily descending terrain (orange: Engine ECU attack run; cyan: baseline run). 
        \textbf{Bottom panel:}
        The fuel efficiency ratio (solid grey line) is the performance in the attack run divided by the performance in the baseline run. The overlaid, blue dashed line, is the fit of the continuous model. The green and red horizontal lines at the top and bottom indicate the times when the bonware and malware (resp.) are active. The model captures the rapid decline to an equilibrium state around 92\% performance as well as the more gradual recovery after the cyber attack.
    }
       \Description{
        Progression of data over time. 
        Top panel. Fuel efficiency of a heavy truck, carrying no cargo, during a run on steadily descending terrain (orange: Engine ECU attack run; cyan: baseline run). 
        Bottom panel.
        The fuel efficiency ratio (solid grey line) is the performance in the attack run divided by the performance in the baseline run. The overlaid, blue dashed line, is the fit of the continuous model. The green and red horizontal lines at the top and bottom indicate the times when the bonware and malware are active. The model captures the rapid decline to an equilibrium state around 92 percent performance as well as the more gradual recovery after the cyber attack.
    }
    \label{fig:model fits}
\end{figure}

Our modeling approach requires one further data processing step, which is illustrated in Fig.~\ref{fig:model fits}. The top panel shows a baseline and attack performance curve. The ratio of these curves (i.e., performance under attack divided by the baseline value) is shown in the bottom panel -- this ratio is close to $1.0$ when performance under cyber attack is similar to the baseline performance, and less than $1.0$ when the cyber attack is detrimental to performance. This performance ratio is the measure of functionality that we use for our modeling. Also in the bottom panel, we show the fitted ``piecewise constant'' model that is described in Subsection~\ref{subsection:piecewise constant model}. The red and green intervals indicate the activity periods of the malware and bonware, respectively. When they are inactive, these effectiveness parameters are $0$, otherwise they are $M$ and $B$ respectively. We can see that the model captures the drop in performance when the malware is active.
To summarize and interpret our data, we applied this model to the data from each experimental condition separately.  In order to automate the parameter estimation process, we implemented our piecewise constant model using a Bayesian inference engine \cite{pyjags,Matzke2017}.\footnote{{JAGS uses Markov chain Monte Carlo (MCMC) methods to estimate parameters. We confirmed convergence of four MCMC chains using a potential scale reduction factor (often called $\hat{R}$; \cite{gelman1995bayesian}) criterion of $1.05$.}}  To further facilitate the automation, we additionally allowed the model to estimate the time points  where performance begins to decline ($t_1$) and recover ($t_2$).
{This implementation lets us quantify the uncertainty in parameter estimates and has the advantage of being fully automatic and easily extendable for future projects.}

As illustrated in Fig.~\ref{fig:model fits}, our mathematical model and the experimental data exhibit a similar pattern and the model produces a good fit. Moreover, the estimates of parameters $M$ and $B$ attain values that reflect the temporal behaviors of experimental data (e.g., high $B$ is associated with rapid recovery, and the ratio of $B/(M+B)$ approximates the performance equilibrium when the cyber attack is active).

\begin{figure}[ht]
    \centering
    \includegraphics[width=0.75\columnwidth,trim=0 18 0 0,clip]{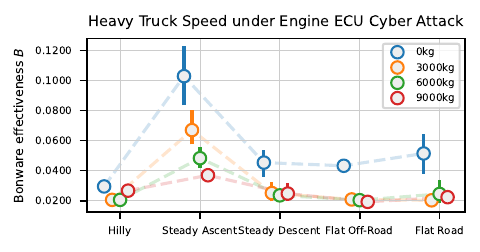}
    \includegraphics[width=0.75\columnwidth,trim=0 0 0 15,clip]{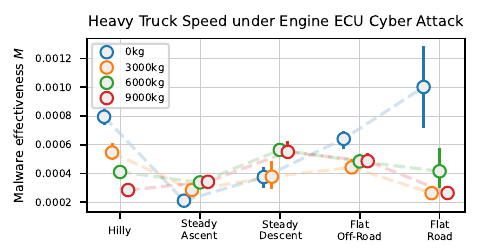}
    \caption{Parameter estimates of the piecewise continuous model applied to performance of a heavy truck under a cyber attack on the engine ECU. The top panel displays the effectiveness of the bonware as a function of terrain and cargo weight, while the bottom panel displays effectiveness of the malware. One observation is that the effectiveness of the bonware is generally much higher than that of the malware, largely due to the physical resilience of the truck machinery. Round markers indicate parameter estimates, the intervals around the markers are 95\% credible intervals. Each panel summarizes 1,200 runs (5 terrains by 4 cargo weights by 30 repetitions, once under baseline and once under cyber attack).}
    \Description{Two panels with multiple line graphs.  The top graph shows the Bonware effectiveness B for a heavy truck's speed under the Engine ECU Cyber Attack.  The bottom graph shows the Malware effectiveness.  On the horizontal axis are different subroutes.  Both graphs have four lines, for different cargo weights.  In the Bonware graphs, all lines peak at the Steady Ascent subroute, and the 0kg cargo condition has the highest effectiveness.  The Malware graph shows a more complex pattern, with different lines peaking at different subroutes and differing orderings of cargo weight conditions.}
    \label{fig:heavy ecu param estimates}
\end{figure}

The modeling approach allows us to summarize complex time series with two interpretable parameters: the malware effectiveness $M$ and the bonware effectiveness $B$. This facilitates comparison of the cyber resilience of our trucks under various conditions. For example, Fig.~\ref{fig:heavy ecu param estimates} shows the pattern of results of an engine ECU cyber attack on a heavy truck (note: higher $B$ means more effective bonware and higher $M$ means more effective malware). At a glance at the top panel, we can determine that the truck resists the cyber attack better when it is not hauling cargo (blue markers are always higher), and especially so when the terrain is a steady ascent (the difference is especially pronounced in that subroute). The bottom panel shows that the cyber attack is relatively more effective when the cargo is light (blue markers are often higher) and especially the road is flat (the blue line has its peak there). Comparing the magnitude of the parameter estimates between the two panels ($B$ is greater than $M$ by at least an order of magnitude) tells us that this cyber attack, even at its most effective, only has a modest effect on functionality.

\subsection{\texorpdfstring{Resilience $R$}{Resilience R}}\label{resilience results}

We will use the experimental data to compute the $R$ statistic introduced in Eq.~(\ref{eq:R}) in subsection~\ref{sec:auc based}.  
The calculation involves (1) finding the area $\ma_\text{attack}$ under the performance curve during the time when the cyber attack is active, then (2) finding the corresponding area $\ma_\text{baseline}$ under the baseline performance curve, and (3) dividing the former by the latter.  If the resulting $R$ is $1.0$, then the cyber attack had no detrimental effect on performance.  $R$ of $0.0$ means that performance was reduced by 100\%.

\begin{figure}[ht]
    \centering
    \includegraphics[width=\columnwidth,bb=0 307 757 537]{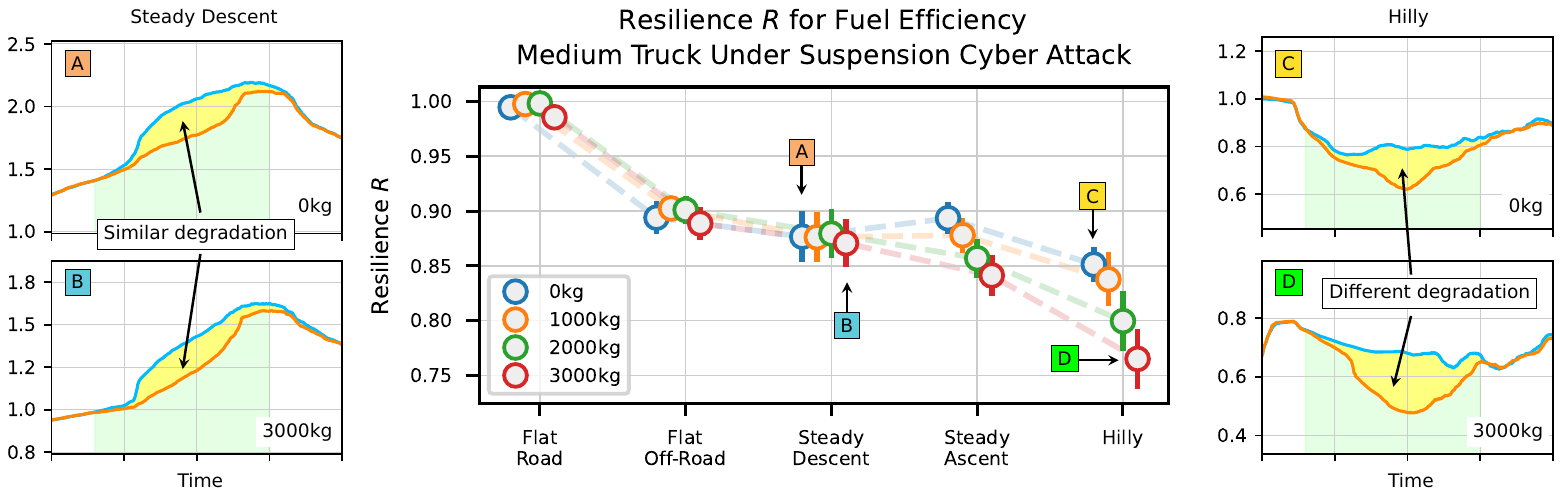}
    \caption{An illustration of the resilience measure $R$. The central panel shows the value of $R$ for 20 different conditions (five subroutes and four cargo weights). Each marker shows $R$ computed using the average of 30 runs of a suspension cyber attack and the average of 30 baseline runs (error bars indicating 95\% confidence intervals), in this case by a medium-weight truck. The panel thus summarizes 1,200 experimental runs. The four side panels illustrate the construction of $R$. Each side panel shows the construction of one marker in the central panel. The blue (upper) curve indicates functionality under the baseline scenario. Note that the baseline functionality differs between subroutes (left vs.\ right panels) and between cargo weight conditions (top vs.\ bottom). The orange (bottom) curve indicates the functionality during the cyber attack. The shaded yellow region between the curves is the effect of the attack: a temporary reduction in functionality. $R$ is the ratio of area under the orange curve to that under the blue curve (Eq.~(\ref{eq:R})) and can be interpreted as a measure of resilience: it is the remaining fraction of functionality while under attack. Those values of $R$ are displayed in the central panel, which shows that resilience is high on flat road, but lower on hilly terrain. It further shows a notable effect of cargo weight in the hilly terrain as well as during the steady ascent, but no effect of cargo weight in the steady descent, flat road, or flat off-road subroutes.}
     \Description{An illustration of the resilience measure R. The central panel shows the value of R for 20 different conditions (five subroutes and four cargo weights). Each marker shows R computed using the average of 30 runs of a suspension cyber attack and the average of 30 baseline runs (error bars indicating 95 percent confidence intervals), in this case by a medium-weight truck. The panel thus summarizes 1,200 experimental runs. The four side panels illustrate the construction of R. Each side panel shows the construction of one marker in the central panel. The blue (upper) curve indicates functionality under the baseline scenario. Note that the baseline functionality differs between subroutes (left vs.\ right panels) and between cargo weight conditions (top vs.\ bottom). The orange (bottom) curve indicates the functionality during the cyber attack. The shaded yellow region between the curves is the effect of the attack: a temporary reduction in functionality. R is the ratio of area under the orange curve to that under the blue curve and can be interpreted as a measure of resilience: it is the remaining fraction of functionality while under attack. Those values of R are displayed in the central panel, which shows that resilience is high on flat road, but lower on hilly terrain. It further shows a notable effect of cargo weight in the hilly terrain as well as during the steady ascent, but no effect of cargo weight in the steady descent, flat road, or flat off-road subroutes.}
    \label{fig:auc explainer}
\end{figure}

Fig.~\ref{fig:auc explainer} illustrates that the resilience measure $R$ (based on the area under the curve concept) follows our intuitive understanding of what a measure of cyber resilience should do: it is higher if performance is impacted by cyber attack less, and lower if it is impacted more. It is relatively unimpacted in cases where no impact was expected (e.g., on the flat road subroutes in Fig.~\ref{fig:auc explainer}, there is no difference between cargo weight conditions), and it gives orderly results when differences are expected (e.g., when in Fig.~\ref{fig:auc explainer} the $R$ is affected by cargo weights, it is consistently lower for heavier cargo).

\begin{figure}[ht]
    \centering
    \includegraphics[width=0.75\columnwidth,trim=0 0 0 0,clip]{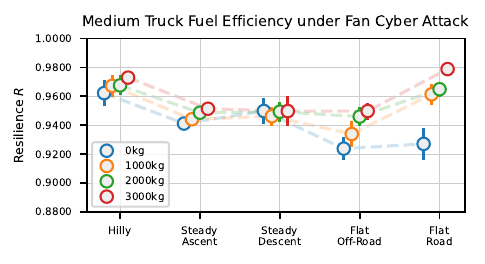}
    \caption{The $R$ measure for a medium truck subjected to a fan cyber attack. Round markers indicate parameter estimates, the intervals around the markers are 95\% credible intervals. Each panel summarizes 1,200 runs (5 terrains by 4 cargo weights by 30 repetitions, once under baseline and once under cyber attack).}
    \Description{A line graph showing the Resilience R of a medium truck's fuel efficiency under the Fan Cyber Attack, plotted over different subroutes, for different cargo weight conditions. The higher cargo weight conditions tend to have higher resilience scores, and the lowest resilience is seen in the Flat Off-Road and Flat Road conditions.}
    \label{fig:interaction plot}
\end{figure}

Fig.~\ref{fig:interaction plot} shows results for the same truck when it is subjected to the cyber attack on the engine fan controller. Here, we see a different pattern of results, with the effect generally being greater when the cargo is lighter. {This, too, accords with our intuitions about resilience} {and the nature of the attack, which ultimately forces the truck to require more engine power to maintain vehicle speed}: {In a baseline run, } {with lighter cargo and flatter terrain, the truck requires comparatively less engine power to maintain vehicle speed} {than with heavier cargo and difficult terrain.} {Consequently,} {the difference in required engine power between corresponding baseline and fan attack runs will be greater and more pronounce on lighter cargo and flatter terrain than with heavier cargo and more challenging terrain. Regardless,} the results remain ordered and show a great deal of consistency. There seems to be much less difference between subroutes during this attack. Also, on average, the loss of functionality due to this attack is smaller than that due to the attack on the engine ECU.

Taken together, our experimental data support the validity of the $R$ measure as a quantitative measure of cyber resilience.

\section{Conclusions and Future Work}\label{sec:conclusions}

We have reported results of the \textit{Quantitative Measurement of Cyber Resilience} project, in which we obtain experimental data with physical-digital twins of several cargo trucks and analyze the data with mathematical modeling of time series in order to quantify and measure the cyber resilience of the trucks.  We were successful in generating data with apparent fidelity, showing that changes in the setup of the experimental runs (e.g., heavier cargo, more challenging terrain) result in differences in performance that accord with our subject matter expertise as well as common-sense expectations.

We proposed two types of summary statistics. One is a measure of resilience based on the area under the performance curve. Another type is based on fits of a mathematical model to temporal evolution of the performance curve, and measures the effectiveness of malware and bonware. These measures seem to capture the salient patterns in the experimental data succinctly, supporting their use as quantitative measures of cyber resilience.

{Based on the results reported in this paper, we have formulated a proposed practical method for measuring cyber resilience of a system under test. The details of the method go beyond the scope of this paper, but are described in a separate publication \cite{kottmethodology}.}

We believe there is much that could still be learned with our (or a similar) test bed. Data is relatively easy and fast to gather, and more variables can still be introduced. Additionally, similar test beds could be constructed for other types of vehicles, but also for other diverse types of complex equipment, infrastructure, or critical digital services. For the trucks, the test bed could still be augmented with additional derived measures of functionality, such as maneuverability.

Similarly, there are further potential developments in the mathematical modeling aspect. New models could be implemented to allow for multivariate functionality, to account for trade-offs between different performance parameters. 

{Besides the practical significance of such measures, our research makes broader theoretical contributions. We offer an explicit conceptualization of cyber resilience as a characteristic of the process -- the ``battle'' between malware and bonware of a system under consideration. We further offer that this battle might be usefully modeled in a system-dynamic paradigm and illustrate this perspective with a specific class of models. Furthermore, we introduce notions of malware effectiveness and bonware effectiveness and of their temporal dynamics as key determinants of cyber resilience.}

\appendices
\section{Stability}
\label{appendix:A}

A linear differential equation is said to be stable, in the
sense of Laplace \cite{struble, teschl2012ordinary}, if all solutions are bounded as $t\to\infty.$  We will show that Equation (\ref{eq:00}) is stable due to the non-negativity of both $\malware(t)$ and $\bonware(t)$.  Our approach is to consider the simplified cases where each of $\malware(t)$ and $\bonware(t)$ are zero, show that each of these equations is stable, and then show that this implies Equation (\ref{eq:00}) is stable.

Consider Equation (\ref{eq:00}) with $\bonware(t)=0.$

\begin{equation}
    \frac{d\functionality}{dt} + \malware(t) \functionality(t) = 0.
    \label{eq:11}
\end{equation}
The solution is 
\begin{equation}
  \functionality(t) = F_0 e^{-\int_0^t M(p)\, dp},
    \label{eq:22}
\end{equation}
and since $\malware(t) \ge 0$,  $\functionality(t)$ is bounded everywhere ($0  \le \functionality(t) \le F(0)$).

Let us now discuss Equation (\ref{eq:00}) with $\malware(t)=0.$

Then 

\begin{equation}
    \frac{d\functionality}{dt} + \bonware(t)(\functionality(t) - \Fnominal) =0.
    \label{eq:001}
\end{equation}
or
\begin{equation}
    \frac{d\Phi}{dt} + \bonware(t)\Phi(t) =0. \quad \text{where } \Phi(t) = F(t) - \Fnominal
    \label{eq:002}
\end{equation}

Equation (\ref{eq:002}) has solution
\begin{equation*}
    \Phi(t)  = (F(0)-\Fnominal) e^{-  \int_0^t B(p) \, dp }.
\end{equation*}
so that 
\begin{equation*}
    \functionality(t)  = \Fnominal+ (F(0)-\Fnominal) e^{-  \int_0^t B(p) \, dp }.
\end{equation*}
Since we require $\bonware(t) \ge 0$, $\functionality(t)$ is bounded
everywhere ($F(0) \le F(t) \le \Fnominal$).

We've shown that in intervals when either $\malware(t)$ or $\bonware(t)$ are
zero, the solution to the differential equation is bounded.  Now
consider the case where both $\malware(t)$ and $\bonware(t)$ are strictly
positive on an interval $(t_1,t_2)$, initially with $F(t_1)>0$  Let $\functionality(t)$ be a solution to
(\ref{eq:00}) and let $\functionality_1(t)$ be a solution to (\ref{eq:001}).
Subtracting, we obtain:
\begin{equation}
  \frac{d}{dt} (\functionality_1(t)-\functionality(t)  )
  + \bonware (t) (\functionality_1(t)-\functionality(t)  ) =
  \malware(t) \functionality(t).
    \label{eq:005}
\end{equation}
Assume that there exists a time that $\functionality(t)=0.$  Then by
(\ref{eq:005}), 
\begin{equation*}
\frac{d \functionality_1(t) }{dt} + \bonware(t) \functionality_1(t) =0.
\end{equation*}
But $\functionality_1(t)$ satisfies equation (\ref{eq:001}).  So
$\bonware(t)$ must be zero contradicting our assumption that both
$\malware(t)$ and $\bonware(t)$ are positive.  Thus
$\functionality(t)>0$ on $(t_1,t_2)$ and thus is bounded below.

Assuming again that $\functionality(t)$ satisfies (\ref{eq:00})
and $\functionality_1(t)$ satisfies (\ref{eq:001}).  We further assume,
$\functionality_1(t_1)=\functionality(t_1).$  Then at $t_1$, we have
$\frac{d\functionality(t)}{dt} < \frac{d\functionality_1(t)}{dt}$
implying that at time $t^\star>t_1$,
$\functionality(t^\star)<\functionality_1(t^\star)$.  So $\Fnominal \ge
F_1(t) \ge F(t)$ everywhere, and $\functionality(t)$ is bounded above
by $\Fnominal.$  We have thus shown that if $\functionality(t)$ is a
solution to (\ref{eq:00}), then $\Fnominal \ge \functionality(t) \ge
0$ and thus Equation (\ref{eq:00}) is stable in the sense of Laplace. 

\section{Obtaining the solution to the LTV model}
\label{appendix:B}
\label{sec:B}
 Here, we develop the solution (Equation
(\ref{eq:6})), to the LTV model, (Equation \ref{eq:linear:model})), in
Section \ref{sec:linear}.  We start with Equation~\ref{eq:linear:model}: 
\begin{equation*}
 \frac{dF}{dt}+(\lambda - \omega t) F(t) = \Fnominal (\alpha - \beta t).
\end{equation*}
Multiplying both sides of this equation by the integrating factor $\omega(t) =
e^{\lambda t - \frac{\omega t^2}{2}}$ and combining terms, we obtain:
\begin{equation*}
\frac{d}{dt} \left( \frac{F}{\Fnominal} e^{\lambda t - \frac{\omega t^2}{2}}
      \right)=(\alpha-\beta t)  e^{\lambda t - \frac{\omega t^2}{2}}.
    \end{equation*}
    Integrating both sides of this equation with respect to $t$:
\begin{equation*}
 \frac{F(t)}{\Fnominal}  e^{\lambda t - \frac{\omega t^2}{2}} -
      \frac{F(0)}{\Fnominal}=\int_0^t (\alpha-\beta \tau) e^{\lambda
        \tau - \frac{\omega \tau^2}{2}} d\tau.
    \end{equation*}
    Completing the square in the exponent:
 \begin{equation*}
  \frac{F(t)}{\Fnominal}       = e^{\frac{\omega t^2}{2} - \lambda t}
      \left(
        \frac{F(0)}{\Fnominal} + e^{\frac{\lambda^2}{2\omega}}
        \int_0^t
        (\alpha-\beta\tau) e^{-(\frac{\lambda}{\sqrt{2\omega}} - \sqrt{\frac{\omega}{2}}\tau)^2 }
        d\tau
      \right).
\end{equation*}
With the change of variables setting the quadratic in the exponent to
$p^2$:
\begin{equation*}
  \frac{F(t)}{\Fnominal}       = e^{\frac{\omega t^2}{2} - \lambda t}
      \left(
        \frac{F(0)}{\Fnominal} + \sqrt{\frac{2}{\omega}}e^{\frac{\lambda^2}{2\omega}}
        \int_{-\frac{\lambda}{\sqrt{2\omega}}}^{\sqrt{\frac{\omega}{2}}t - \frac{\lambda}{\sqrt{2\omega}}}     
        \left(\alpha- \frac{\beta\lambda}{\omega} -
          \beta\sqrt{\frac{2}{\omega}}p \right) e^{-p^2 } d p
      \right).
\end{equation*}
The second term of the integral can be integrated immediately, and the
first term can be written in terms of the error function.  With
$\erf(z) = \frac{2}{\sqrt{\pi}} \int_0^z e^{-\tau^2} d\tau,$ the result is
  \begin{equation*}
    \frac{F(t)}{\Fnominal}=
e^{\frac{\omega t^2  }{2}-\lambda  t} \left(\frac{F(0)}{\Fnominal}+\frac{\sqrt{\frac{\pi }{2}} e^{\frac{\lambda
   ^2}{2 \omega }} (\alpha  \omega -\beta  \lambda ) \left(\text{erf}\left(\frac{\lambda
   }{\sqrt{2 \omega }}\right)+\text{erf}\left(\frac{t \omega
     -\lambda }{\sqrt{2 \omega }}\right)\right)}{\omega ^{3/2}}+\frac{\beta 
   \left(e^{\lambda  t-\frac{\omega t^2 }{2}}-1\right)}{\omega }\right).
\end{equation*}

\small
\bibliographystyle{IEEEtran}
\IEEEtriggeratref{22}
\bibliography{joint.bib}\null

\begin{thebibliography}{10}
\providecommand{\url}[1]{#1}
\csname url@samestyle\endcsname
\providecommand{\newblock}{\relax}
\providecommand{\bibinfo}[2]{#2}
\providecommand{\BIBentrySTDinterwordspacing}{\spaceskip=0pt\relax}
\providecommand{\BIBentryALTinterwordstretchfactor}{4}
\providecommand{\BIBentryALTinterwordspacing}{\spaceskip=\fontdimen2\font plus
\BIBentryALTinterwordstretchfactor\fontdimen3\font minus
  \fontdimen4\font\relax}
\providecommand{\BIBforeignlanguage}[2]{{%
\expandafter\ifx\csname l@#1\endcsname\relax
\typeout{** WARNING: IEEEtran.bst: No hyphenation pattern has been}%
\typeout{** loaded for the language `#1'. Using the pattern for}%
\typeout{** the default language instead.}%
\else
\language=\csname l@#1\endcsname
\fi
#2}}
\providecommand{\BIBdecl}{\relax}
\BIBdecl

\bibitem{kott2022}
A.~Kott, M.~Weisman, and J.~Vandekerckhove, ``Mathematical modeling of cyber
  resilience,'' \emph{Proceedings of IEEE Military Communications Conference},
  pp. 835--840, Dec. 2022.

\bibitem{weisman23}
M.~Weisman, A.~Kott, and J.~Vandekerckhove, ``Piecewise linear and stochastic
  models for the analysis of cyber resilience,'' \emph{57th Annual Conference
  on Information Sciences and Systems}, Mar. 2023.

\bibitem{kott2019cyber}
A.~Kott and I.~Linkov, \emph{Cyber resilience of systems and networks}.\hskip
  1em plus 0.5em minus 0.4em\relax New York, NY: Springer International
  Publishing, 2019.

\bibitem{smith23towards}
S.~Smith, ``Towards a scientific definition of cyber resilience,'' in
  \emph{18th International Conference on Cyber Warfare and Security ({ICCWS
  2023})}.\hskip 1em plus 0.5em minus 0.4em\relax Red Hook, NY: Academic
  Conferences Ltd, Mar.9--10 2023, pp. 1--9.

\bibitem{linkov2018risk}
I.~Linkov, B.~D. Trump, and J.~Keisler, ``Risk and resilience must be
  independently managed,'' \emph{Nature}, vol. 555, p. 7694, 2018.

\bibitem{bozdal2018august}
M.~Bozdal, M.~Samie, and I.~Jennions, ``A survey on {CAN} bus protocol:
  Attacks, challenges, and potential solutions.'' in \emph{2018 International
  Conference on Computing, Electronics \& Communications Engineering}.\hskip
  1em plus 0.5em minus 0.4em\relax Ieee, August 2018, pp. 201--205.

\bibitem{Pang2004}
H.~Pang and C.~Brace, ``Review of engine cooling technologies for modern
  engines,'' \emph{Proceedings of the Institution of Mechanical Engineers, Part
  D: Journal of Automobile Engineering}, vol.~11, no. 218, pp. 1209--1215,
  2004.

\bibitem{kott2018}
A.~Kott, P.~Th\'{e}ron, M.~Dra\v{s}ar, E.~Dushku, B.~LeBlanc, P.~Losiewicz,
  ..., and K.~Rzadca, ``Autonomous intelligent cyber-defense agent (aica)
  reference architecture,'' 2018, arXiv:1803.10664.

\bibitem{kott2021cyber}
A.~Kott, M.~S. Golan, B.~D. Trump, and I.~Linkov, ``Cyber resilience: by design
  or by intervention?'' \emph{Computer}, vol.~54, no.~8, pp. 112--117, 2021.

\bibitem{kott2020doers}
A.~Kott and P.~Theron, ``Doers, not watchers: Intelligent autonomous agents are
  a path to cyber resilience,'' \emph{IEEE Security \& Privacy}, vol.~18,
  no.~3, pp. 62--66, 2020.

\bibitem{dillon1947resilience}
J.~Dillon, ``Resilience of fibers and fabrics,'' \emph{Textile Research
  Journal}, vol.~17, no.~4, pp. 207--213, 1947.

\bibitem{hoffman1948generalized}
R.~Hoffman, ``A generalized concept of resilience,'' \emph{Textile Research
  Journal}, vol.~18, no.~3, pp. 141--148, 1948.

\bibitem{herrman2011resilience}
H.~Herrman, D.~Stewart, N.~Diaz-Granados, E.~Berger, B.~Jackson, and T.~Yuen,
  ``What is resilience?'' \emph{The Canadian Journal of Psychiatry}, vol.~56,
  no.~5, pp. 258--265, 2011.

\bibitem{wu2013understanding}
G.~Wu, A.~Feder, H.~Cohen, J.~Kim, S.~Calderon, D.~Charney, and A.~Math{\'e},
  ``Understanding resilience,'' \emph{Frontiers in behavioral neuroscience},
  vol.~7, p.~10, 2013.

\bibitem{folke2010resilience}
C.~Folke, S.~Carpenter, B.~Walker, M.~Scheffer, T.~Chapin, and
  J.~Rockstr{\"o}m, ``Resilience thinking: integrating resilience, adaptability
  and transformability,'' \emph{Ecology and society}, vol.~15, no.~4, 2010.

\bibitem{dupont2019cyber}
B.~Dupont, ``The cyber-resilience of financial institutions: significance and
  applicability,'' \emph{Journal of cybersecurity}, vol.~5, no.~1, p. tyz013,
  2019.

\bibitem{hausken2020cyber}
K.~Hausken, ``Cyber resilience in firms, organizations and societies,''
  \emph{Internet of Things}, vol.~11, p. 100204, 2020.

\bibitem{herrington2013future}
L.~Herrington and R.~Aldrich, ``The future of cyber-resilience in an age of
  global complexity,'' \emph{Politics}, vol.~33, no.~4, pp. 299--310, 2013.

\bibitem{huang2022reinforcement}
Y.~Huang, L.~Huang, and Q.~Zhu, ``Reinforcement learning for feedback-enabled
  cyber resilience,'' \emph{Annual reviews in control}, vol.~53, pp. 273--295,
  2022.

\bibitem{jensen2015challenges}
L.~Jensen, ``Challenges in maritime cyber-resilience,'' \emph{Technology
  Innovation Management Review}, vol.~5, no.~4, p.~35, 2015.

\bibitem{zou2021cyber}
B.~Zou, P.~Choobchian, and J.~Rozenberg, ``Cyber resilience of autonomous
  mobility systems: cyber-attacks and resilience-enhancing strategies,''
  \emph{Journal of transportation security}, pp. 1--19, 2021.

\bibitem{alexeev2017constructing}
A.~Alexeev, D.~Henshel, K.~Levitt, P.~McDaniel, B.~Rivera, S.~Templeton, and
  M.~Weisman, ``Constructing a science of cyber-resilience for military
  systems,'' in \emph{NATO IST-153 Workshop on Cyber Resilience}, 2017, pp.
  23--25.

\bibitem{henshel}
D.~S. Henshel, K.~Levitt, S.~Templeton, M.~G. Cains, A.~Alexeev, B.~Blakely,
  P.~McDaniel, G.~Wehner, J.~Rowell, and M.~Weisman, ``The science of cyber
  resilience: Characteristics and initial system taxonomy,'' \emph{Fifth World
  Conference on Risk}, 2019.

\bibitem{ligo2021how}
A.~K. Ligo, A.~Kott, and I.~Linkov, ``How to measure cyber-resilience of a
  system with autonomous agents: Approaches and challenges,'' \emph{IEEE
  Engineering Management Review}, vol.~49, no.~2, pp. 89--97, 2021.

\bibitem{linkov2013resilience}
I.~Linkov, D.~Eisenberg, K.~Plourde, T.~Seager, J.~Allen, and A.~Kott,
  ``Resilience metrics for cyber systems,'' \emph{Environment Systems and
  Decisions}, vol.~33, no.~4, pp. 471--476, 2013.

\bibitem{beling2021developmental}
P.~Beling, B.~Horowitz, and T.~McDermott, ``Developmental test and evaluation
  {(DTE\&A)} and cyberattack resilient systems,'' Systems Engineering Research
  Center, Hoboken, NJ, Technical Report Serc-2021-tr-015, September 2021.

\bibitem{hosseini2016review}
S.~Hosseini, K.~Barker, and J.~Ramirez-Marquez, ``A review of definitions and
  measures of system resilience,'' \emph{Reliability Engineering \& System
  Safety}, vol. 145, pp. 47--61, 2016.

\bibitem{kott2021to}
A.~Kott and I.~Linkov, ``To improve cyber resilience, measure it,''
  \emph{Computer}, vol.~54, no.~2, pp. 80--85, Feb. 2021.

\bibitem{hoppe2007sniffing}
T.~Hoppe and J.~Dittman, ``Sniffing/replay attacks on {CAN} buses: A simulated
  attack on the electric window lift classified using an adapted cert
  taxonomy,'' in \emph{Proceedings of the 2nd workshop on embedded systems
  security (WESS)}, 2007, pp. 1--6.

\bibitem{koscher2010experimental}
K.~Koscher, A.~Czeskis, F.~Roesner, S.~Patel, T.~Kohno, S.~Checkoway, D.~McCoy,
  B.~Kantor, D.~Anderson, H.~Shacham \emph{et~al.}, ``Experimental security
  analysis of a modern automobile,'' in \emph{2010 IEEE symposium on security
  and privacy}.\hskip 1em plus 0.5em minus 0.4em\relax Ieee, 2010, pp.
  447--462.

\bibitem{miller2013adventures}
C.~Miller and C.~Valasek, ``Adventures in automotive networks and control
  units,'' \emph{Def Con}, vol.~21, no. 260-264, pp. 15--31, 2013.

\bibitem{miller2015remote}
------, ``Remote exploitation of an unaltered passenger vehicle,'' \emph{Black
  Hat USA}, vol. 2015, no. S 91, 2015.

\bibitem{foster2015fast}
\BIBentryALTinterwordspacing
I.~Foster, A.~Prudhomme, K.~Koscher, and S.~Savage, ``Fast and vulnerable: A
  story of telematic failures,'' in \emph{9th USENIX Workshop on Offensive
  Technologies (WOOT 15)}.\hskip 1em plus 0.5em minus 0.4em\relax Washington,
  D.C.: USENIX Association, Aug. 2015. [Online]. Available:
  \url{https://www.usenix.org/conference/woot15/workshop-program/presentation/foster}
\BIBentrySTDinterwordspacing

\bibitem{daily2016towards}
J.~Daily, R.~Gamble, S.~Moffitt, C.~Raines, P.~Harris, J.~Miran, I.~Ray,
  S.~Mukherjee, H.~Shirazi, and J.~Johnson, ``Towards a cyber assurance testbed
  for heavy vehicle electronic controls,'' \emph{SAE International Journal of
  Commercial Vehicles}, vol.~9, no.~2, pp. 339--349, 2016.

\bibitem{bozdal2018hardware}
M.~Bozdal, M.~Randa, M.~Samie, and I.~Jennions, ``Hardware trojan enabled
  denial of service attack on {CAN} bus,'' \emph{Procedia Manufacturing},
  vol.~16, pp. 47--52, 2018.

\bibitem{wang2018delay}
Q.~Wang, Y.~Qian, Z.~Lu, Y.~Shoukry, and G.~Qu, ``A delay based plug-in-monitor
  for intrusion detection in controller area network,'' in \emph{2018 Asian
  Hardware Oriented Security and Trust Symposium (AsianHOST)}, Dec 2018, pp.
  86--91.

\bibitem{shikata2019digital}
H.~Shikata, T.~Yamashita, K.~Arai, T.~Nakano, K.~Hatanaka, and H.~Fujikawa,
  ``Digital twin environment to integrate vehicle simulation and physical
  verification,'' \emph{SEI Technical Review}, vol.~88, pp. 18--21, 2019.

\bibitem{kottmethodology}
A.~Kott, M.~Weisman, J.~Vandekerckhove, J.~Ellis, T.~Parker, B.~Murphy, and
  S.~Smith, ``A methodology for quantitative measurement of cyber resilience
  (qmocr),'' \emph{ARL-TR-9672}, 2023.

\bibitem{lin}
H.~Lin and P.~Antsaklis, ``Stability and stabilizability of switched linear
  systems: A short survey of recent results,'' in \emph{Proceedings of the 2005
  IEEE International Symposium on, Mediterrean Conference on Control and
  Automation Intelligent Control, 2005.}, 2005, pp. 24--29.

\bibitem{ellis2022experimental}
J.~Ellis, T.~Parker, J.~Vandekerckhove, B.~Murphy, S.~Smith, A.~Kott, and
  M.~Weisman, ``Experimental infrastructure for study of measurements of
  resilience,'' \emph{Proceedings of IEEE Military Communications Conference},
  pp. 841--846, Dec. 2022.

\bibitem{toyama2018pasta}
T.~Toyama, T.~Yoshida, H.~Oguma, and T.~Matsumoto, ``{PASTA}: Portable
  automotive security testbed with adaptability,'' \emph{Black Hat Europe
  2018}, Dec. 3--6, 2018.

\bibitem{unity2021unity}
\BIBentryALTinterwordspacing
{Unity Technologies}. (2021) The leading platform for creating interactive,
  real-time content. (v2020.3.18f1). [Online]. Available:
  \url{https://unity.com/}
\BIBentrySTDinterwordspacing

\bibitem{opentap2021whitepaper}
\BIBentryALTinterwordspacing
{Keysight Technologies}. (2021) ``{O}pen source in test automation''. [Online].
  Available:
  \url{https://opentap.io/assets/Open-Source-in-Test-Automation-v3.pdf/}
\BIBentrySTDinterwordspacing

\bibitem{pyjags}
T.~Miasko, ``pyjags (version 1.3.8),'' 2017, [computer software] (available at:
  https://github.com/tmiasko/pyjags.

\bibitem{Matzke2017}
\BIBentryALTinterwordspacing
D.~Matzke, U.~Boehm, and J.~Vandekerckhove, ``Bayesian inference for
  psychology, part {III}: Parameter estimation in nonstandard models,''
  \emph{Psychonomic Bulletin \& Review}, vol.~25, no.~1, pp. 77--101, Nov.
  2017. [Online]. Available: \url{https://doi.org/10.3758/s13423-017-1394-5}
\BIBentrySTDinterwordspacing

\bibitem{gelman1995bayesian}
A.~Gelman, J.~Carlin, H.~Stern, and D.~Rubin, \emph{Bayesian data
  analysis}.\hskip 1em plus 0.5em minus 0.4em\relax Chapman and Hall/CRC, 1995.

\bibitem{struble}
R.~Struble, \emph{Nonlinear differential equations}.\hskip 1em plus 0.5em minus
  0.4em\relax McGraw-Hill, 1962.

\bibitem{teschl2012ordinary}
G.~Teschl, \emph{Ordinary Differential Equations and Dynamical Systems}, ser.
  Graduate studies in mathematics.\hskip 1em plus 0.5em minus 0.4em\relax
  American Mathematical Society, 2012.

\end{thebibliography}
\begin{acronym}[AutoCRATmm]
  \acro{ADF}[ADF]{Active Defense Framework}
  \acro{AUC}[AUC]{area under the curve}
  \acro{AutoCRAT}[AutoCRAT]{
    Automated Cyber Resilience Assessment Tool
  }
  \acro{AWD}[AWD]{all-wheel drive}
  \acro{BSFC}[BSFC]{brake-specific fuel consumption}
  \acro{CA}[CA]{Collaborative Agreement}
  \acro{CAN}[CAN]{Controller Area Network}
  \acro{CGW}[CGW]{central gateway}
  \acro{DIR}[DIR]{detect, isolate, and recover}
  \acro{DoD}[DoD]{Department of Defense}
  \acro{ECU}[ECU]{electronic control unit}
  \acro{IDE}[IDE]{integrated development environment}
  \acro{IP}[IP]{Internet Protocol}
  \acro{KPP}[KPP]{key performance parameter}
  \acro{LDI}[LDI]{livelihood diversification index}
  \acro{MEF}[MEF]{mission essential function}
  \acro{MGV}[MGV]{military ground vehicle}
  \acro{MQTT}[MQTT]{Message Queue Telemetry Transport}
  \acro{OBD2}[OBD2]{on-board diagnostics 2}
  \acro{PASTA}[PASTA]{%
    Toyota Portable Automotive Security Testbed with
    Adaptability
  }
  \acro{PGN}[PGN]{parameter group number}
  \acro{PID}[PID]{proportional-integral-derivative}
  \acro{POI}[POI]{point of interest}
  \acro{PSU/ARL}[PSU/ARL]{
    Penn State University Applied Research Laboratory
  }
  \acro{QMoCR}[QMoCR]{
    Quantitative Measurement of Cyber Resilience
  }
  \acro {RI}[RI]{resilience index}
  \acro {RPM}[RPM]{revolutions per minute}
  \acro {SAE}[SAE]{Society of Automotive Engineers}
  \acro {SLCAN}[SLCAN]{CAN over Serial}
  \acro {SPN}[SPN]{suspect parameter number}
  \acro {USB}[USB]{universal serial bus}
  \acro {CTIS}[CTIS]{Central Tire Inflation System}
  \acro{JLTV}[JLTV]{Joint Light Tactical Vehicle}
  \acro{FMTV}[FMTV]{Family of Medium Tactical Vehicles}
  \acro{HEMTT}[HEMTT]{Heavy Expanded Mobility Tactical Truck}
  \acro{GV}[GV]{Ground Vehicle}
\end{acronym}
\end{document}